\documentclass[10pt,conference,letterpaper]{IEEEtran}
\usepackage{times,amsmath,epsfig}
\usepackage{epsfig,tabularx,amssymb,amsmath,subfigure,multirow, algorithm, algorithmic,graphicx}
\usepackage{pifont}
\usepackage{cases}
\usepackage{subfigure}

\usepackage{xspace}

\title{C-DLSI: An Extended LSI Tailored for Federated Text Retrieval}
\author{%
{Qijun Zhu{\small $~^{1}$}, Dandan Li{\small $~^{2}$}, Dik Lun Lee{\small $~^{3}$} }%
\vspace{1.6mm}\\
\fontsize{10}{10}\selectfont\itshape
Computer Science and Engineering, Hong Kong University of Science and Technology\\
Clear Water Bay, Hong Kong\\
\fontsize{9}{9}\selectfont\ttfamily\upshape
$~^{1}$qijunzhu@cse.ust.hk\\
$~^{2}$ddli@cse.ust.hk\\
$~^{2}$dlee@cse.ust.hk
}
\begin{document}
\maketitle
\begin{abstract}
As the web expands in data volume and in geographical
distribution, centralized search methods become inefficient, leading
to increasing interest in cooperative information retrieval, e.g.,
federated text retrieval (FTR). Different from existing centralized
information retrieval (IR) methods, in which search is done on a
logically centralized document collection,
FTR is composed of a number of peers, each of which is a complete search
engine by itself. To process a query, FTR requires firstly the identification
of promising peers that host the relevant documents and
secondly the retrieval of the most relevant documents from the
selected peers. Most of the existing methods only apply traditional
IR techniques that treat each text collection as a single large
document and utilize term matching to rank the collections.
In this paper, we formalize the problem and identify the properties of FTR,
and analyze the feasibility of extending LSI with clustering to adapt to
FTR, based on which a novel approach called Cluster-based
Distributed Latent Semantic Indexing (C-DLSI) is proposed. C-DLSI
distinguishes the topics of a peer with clustering, captures the
local LSI spaces within the clusters, and consider the relations
among these LSI spaces, thus providing more precise characterization
of the peer. Accordingly, novel descriptors of the peers and a
compatible local text retrieval are proposed. The experimental
results show that C-DLSI outperforms existing methods.
\end{abstract}

%

\section{Introduction}
Due to the highly dynamic nature of the World Wide Web, traditional
search engines (SEs) must face great challenges on scalability and
adaptability. Because of the limited resources available to a search
engine, it is hard to catch up with the fast expansion of the Web
and the frequent updates of its contents. Consequently, the overall
coverage of the search engines with respect to the size of the
entire web deceases with time. We need a scalable and highly
efficient search and index mechanism to make the data on the web in
a timely manner accessible.

To overcome these difficulties, in the past decade, various
information retrieval (IR) methods based on parallel and distributed
computing have been proposed. Among these methods, parallel
information retrieval \cite{Tomasic} that maintains a single index
and employs a server cluster to balance the load has been well
studied and successfully applied in real-world search engines such
as Google. However, it is not scalable with respect to the size and
dynamics of the Web. Furthermore, it cannot handle the hidden deep
web because of privacy issues. To alleviate these problems,
federated information retrieval \cite{Shokouhi} and meta-search
\cite{Meng} were proposed. They send a query simultaneously to
multiple search engines, collect the results from each search engine
after the query has been evaluated separately, and last merge the
results together (i.e., re-ranking). In this way, there is no need
to access directly to the pages or the index at each search engine.
FIR makes it possible to take advantage of the power of different
search engines and provide large coverage of the Web. Since FIR
facilitates cooperation among search engines, it can be more
efficient and effective than meta-search. For this reason, FIR has
attracted much attention in recent years.

As a promising solution to the scalability and adaptability
problems, FIR aims to support search on a large amount of data in a
distributed and self-organizing manner. In the FIR framework, each
search peer indexes and maintains its own document collection, thus
avoiding management problems associated with large data centers. A
broker is introduced to maintain a directory of the peers together
with summarization information, named descriptors, about them. For
query processing, the broker will select peers that have high
potential to return relevant documents for the query according to
the peer descriptors. Note that the broker does not have to know the
peers' indexes or original document sets. In this paper, we only
consider textual documents and content relevance in retrieval, so we
name it federated text retrieval (FTR).

In conventional centralized IR methods, query processing only
focuses on the problem of finding relevant documents using a single
index. On the contrary, FTR requires a three-phase query processing
procedure. First, it identifies promising peers which may return the
most relevant documents. Then it submits the query to the selected
search engines, each of which retrieves the results from its
collection. Finally, it merges the results together and returns them
to the user. Peer selection plays a key role in FTR, which is also
the major concern of this paper. With peer selection, we can make
query evaluation more efficient and, at the same time, save a lot of
computing resources (e.g., power, communication bandwidth, CPU time,
etc.). A number of peer selection approaches \cite{Callan,
Gravano:1} have been proposed, but they are mostly based on the word
histogram of the peers and traditional term matching techniques.

Obviously, the content structure in a collection is significantly
different from that in a document, because a document often focuses
on one topic while a collection may have documents belonging to
different topics. To precisely characterize a heterogeneous
collection, it is necessary to divide it into smaller but more
homogeneous clusters. Inspired by this basic idea, some approaches
\cite{Xu, Shen} utilize clustering to partition the document
collection into different topics and then rank the peers based on
the clusters. Experiments showed that topic-based ranking methods
can substantially improve the quality of peer selection. However,
these studies were only based on heuristics without much rationale
behind them. Besides, they rely heavily on the cluster quality and
ignore the relations among the clusters. In this paper, we propose a
novel approach called Cluster-based Distributed Latent Semantic
Indexing (C-DLSI) based on a formal analysis of the problem. In
particular, C-DLSI, by applying clustering to distinguish the topics
of a peer, extends the traditional LSI scheme and captures delicate
semantic features of the peer, thus providing more precise
characterization of the peer. Moreover, our method is quite scalable
and cost-efficient for the updates.

We detail our main contributions as follows:
\begin{enumerate}
  \item An LSI-based framework (C-DLSI)
for text retrieval in distributed environments was proposed. It
encompasses directory maintenance and query processing.
  \item Identification of the properties of FTR and the feasibility
analysis of extending LSI with clustering to improve the quality of
peer representation. Specifically, the relations among the clusters
are considered in C-DLSI to adapt to the properties of FTR. Our
method is efficient and adapts to frequent collection updates since
only the clusters affected by the updates need to be reindexed.
  \item Based on the analysis of C-DLSI, novel descriptors of the peers
are proposed and a complete federated query processing strategy in
FTR is developed.
  \item The extensive performance evaluation of C-DLSI on a TREC
  dataset. Impacts of different parameter selections are fully
  discussed.
\end{enumerate}

The rest of the paper is organized as follows. In Section 2, we
review the related work on FIR and peer selection. Some bases of our
method, including the framework of FTR, Latent Semantic Indexing
(LSI) and K-means Clustering are introduced in Section 3. In Section
4, we formalize the problem and present our approach C-DLSI in
details. The experimental setup and corresponding results are showed
in Section 5. The last section summarizes the results obtained in
this paper.

\section{Related Work}
Peer selection is a critical problem in FTR and distributed
information retrieval systems in general. It has been studied for
more than a decade. Many methods have been proposed to address this
issue. gGloss (generalized Glossary-Of-Servers Server)
\cite{Gravano:1, Gravano:2} is a well-known method. It keeps
statistics (document frequencies and total weights) about the
servers to estimate which servers are potentially most useful for a
given query. In particular, gGloss(0), a special form of gGloss,
which aggregates all similarity values between the documents and the
query, was shown to be the best and has been widely employed for
comparison \cite{Craswell, Shen, Sogrin}. In this paper, we also use
it as a baseline.

The Collection Retrieval Inference Network (CORI) \cite{Callan,
Powell} is another important work. It drew analogy between
collection ranking and document ranking and applied some form of
$TF\times IDF$ ranking strategy to rank the collections.
Specifically, it replaces TF with DF (document frequency), and IDF
with ICF (inverse collection frequency), the inverse of the
proportion of the collections carrying at least one document which
contains some query terms. Moreover, Yuwono and Lee \cite{Yuwono}
proposed the cue-validity variance (CVV) method for collection
selection. CVV measures the skewness of the distribution of a term
across the collections and estimates the usefulness of the term for
distinguishing a collection from another. Then terms with larger
variances will be given larger weights in index collection ranking.
An evaluation of a number of collection selection methods in a Web
environment was given in \cite{Craswell}. None of these methods
consider the topic space of the peers and utilize semantic
information beyond simple term matching to make a selection.

Latent Semantic Indexing (LSI) \cite{Deerwester} was originally
proposed to take advantage of implicit high-order structure in the
association between terms with documents, namely, semantic
structure, to improve the retrieval of relevant documents. Much
efforts have been made to improve its performance \cite{Letsche,
Jessup, Husbands} or broaden its applications \cite{Pham}. An
earlier work which tried to utilize LSI to improve the peer
selection of FTR is the latent semantic database selection (LSDS)
\cite{Sogrin}. It simply applied LSI to preprocess the document
collections, and conventional selection methods (e.g., CORI) on the
"cleaned" term/document matrices for ranking the collections.
However, this method did not capture the key properties of FTR and,
moreover, inherited the disadvantages of the conventional methods,
e.g., ignoring the topic space of the peers and the drawbacks
inherited from CORI.

To overcome the deficiencies of traditional methods, cluster-based
approaches were proposed to identify the topic space of the peers.
Document clustering was applied to organize collections around
topics and then language modeling was used to represent the topics
\cite{Xu}. This method allows the right topics to be effectively
identified for a given query. However, this method cannot
distinguish the documents within a topic. Shen and Lee \cite{Shen}
proposed another cluster-based method IS-cluster which utilized
cluster descriptors to rank the servers for meta-search engine. We
also use this method as a candidate for comparison in our
experiments. Term correlation was introduced to further improve
cluster-based methods \cite{Zhao}. However, it did not consider the
compatibility issue in FTR, which means that peer selection should
adapt to the local document ranking functions. Further, it is very
difficult to determine the parameters in the method. In this paper,
we extend LSI with the clustering method to especially adapt to FTR
and achieve better retrieval performances.

Recently, uncooperative federated search systems have been studied.
In this case, collections do not disclose their index statistics to
the broker. The broker has to sample documents from each collection
and uses them for collection selection. ReDDE \cite{Si:1} estimates
the number of relevant documents in collections and uses it to
improve collection selection. Estimation of the needed information
for collection selection from an uncooperative peer was addressed in
\cite{Liu}. \cite{Nottelmann} introduced a decision theoretic
framework (DTF) for collection selection, which tries to minimize
the overall costs of federated search including money, time, and
retrieval quality. Similarly, \cite{Si:2} proposed a unified utility
maximization framework (UUM) for resource selection, which evaluates
queries on sampled index. Furthermore, an enhanced model called RUM
\cite{Si:3} was proposed by considering the search effectiveness of
collections. In general, they do not consider the topic space of the
peers either and ignore the semantics. Thus, C-DLSI can also be
embedded into these methods with slightly change (e.g., applied on
the sampling documents) to adapt to this new scenario.

\section{Preliminaries}
In this section, we introduce some preliminaries which act as the
bases of our C-DLSI method. In particular, we first present the
general FTR framework in Section 3.1. Then we describe two important
techniques Latent Semantic Indexing (LSI) and document clustering in
Section 3.2 and Section 3.3, respectively.

\subsection{FTR Framework}
As a federated information retrieval scheme, FTR provides a loose
cooperation among search peers in which each peer maintains its own
local index and a central broker is employed to coordinate the
cooperative text retrieval. Specifically, each peer has a complete
search engine in itself. That is, it has its own crawler, index and
search component for information gathering, organizing and
retrieving, respectively. Besides, the peers share a common
descriptor publishing scheme to disclose to the broker summaries of
information in their repositories. On the other side, the broker of
the FTR system take charge of the query processing by maintaining a
centralized directory, which holds the descriptors of the peers'
local indexes.

\begin{figure}
\centering \epsfig{file=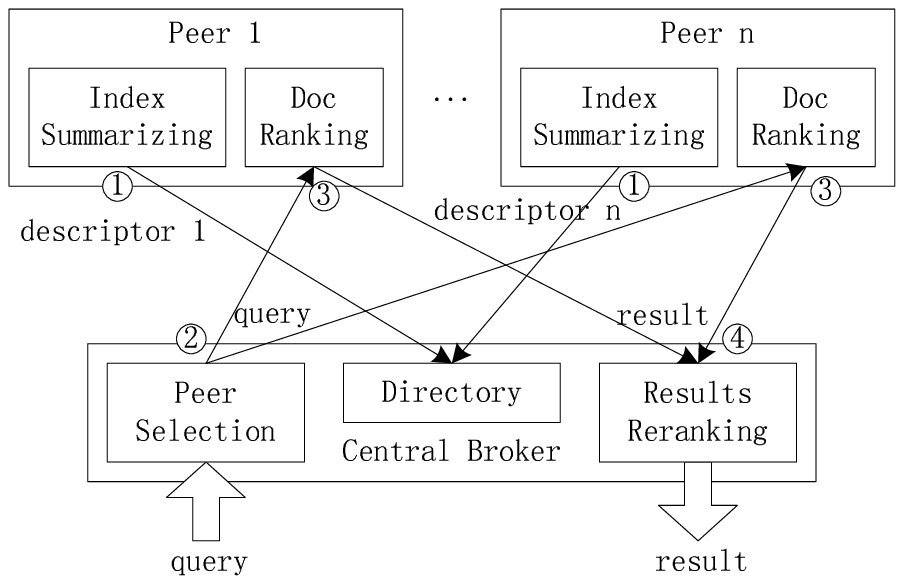, width=3in} \caption{FTR
framework.}
\end{figure}

In FTR, two basic functions are supported: directory maintenance (or
peer descriptor publishing) and query processing. Figure 1 shows the
whole framework of FTR. First, each peer summarizes the descriptor
for its local index and publishes it to the directory in the central
broker. These descriptors are used by the broker to select suitable
peers in query processing. This process is known as the \emph{peer
representation} problem \cite{Bernstein}. Usually, a descriptor
contains connection information together with statistics for each
term in the peer or a limited number of sampled documents. In this
paper, we will provide a novel solution to peer representation in
Section 4.3.

When a query arrives at the broker, the broker selects the most
promising peers which may return the most relevant documents based
on the descriptors. This is the \emph{peer selection} problem
\cite{Callan}. Then the query is forwarded to the selected peers.
Based on the local index, each peer evaluates the query and returns
the results to the broker. Once receiving the results, the broker
will employ a reranking method to properly merge the results
together and present them to the user. It is called the \emph{result
merging} problem. We will consider these issues of the federated
query processing in Section 4.4.

\subsection{Latent Semantic Indexing}
Latent Semantic Indexing (LSI) proposed by Deerwester et al. aims at
taking advantage of the semantic structure of a document collection
to improve retrieval performance. Its objective is to overcome the
fundamental deficiencies of conventional keyword-based information
retrieval techniques. The problem stems from the fact that users are
interested in documents which share the same conceptual content with
the queries, but traditional IR techniques only perform keyword
matching between queries and documents and thus cannot deal with
synonymy and polysemy problems. To bridge the gap, LSI applies
singular value decomposition (SVD) on a term-document matrix to
statistically extract the implicit high-order structure in the
association of terms with documents, which can be used to find the
semantic representations of documents.

LSI is an extension of the vector space model, which approximates
the term-document matrix by the truncated SVD of the matrix. Given
an $m\times n$ term-document matrix $A$, the SVD of $A$ is
\begin{equation} A = U\Sigma V^T,\end{equation}
where $U$ and $V$ have orthonormal columns, $\Sigma$ is a diagonal
matrix having the singular values of $A$ in decreasing order
(denoted as $\sigma_1, \sigma_2, \ldots, \sigma_{rank(A)}$) along
its diagonal, and $T$ denotes the transpose of a vector or a matrix.
LSI decompose $A$ to a lower dimensional vector space $k$ by
retaining only the largest $k$ singular values, where $1\leq k <
rank(A)$. Specifically,
\begin{equation}\label{SVD} A_{k} = U_{k}\Sigma_{k} V_{k}^T,\end{equation}
where $U_{k}$ and $V_{k}$ consist of the first $k$ columns of $U$
and $V$ respectively, and $\Sigma_{k}$ is the $k\times k$ diagonal
matrix containing the largest $k$ singular values of $A$. Because
the number of factors $k$ can be much smaller than the number of
unique terms used to construct this space, terms will not be
independent and the terms with similar meaning will be located near
one another in the LSI space. The relevance score of a document
vector $d$ with a query vector $q$ is measured by the cosine or dot
product between the document vector $d_k$ in LSI space and $q$.
Without loss of generality, we assume that all vectors are
normalized. Then the relevance score can be described as,
\begin{equation} s(d, q) = d^T_k q.\end{equation}
In this paper, we apply the distributed latent semantic indexing to
peer selection and document ranking.

\subsection{Document Clustering}
Since a peer contains a large number of documents, it potentially
covers multiple topics compared to that of a single document. Thus,
a word histogram created for the entire peer cannot provide a
precise description of the peer, making it inadequate to apply a
document ranking approach to peer ranking. A proper peer selection
should consider the topics covered in the peers. In our framework,
we utilize a clustering technique to group the documents of a peer
and treat each cluster as an approximate topic. Then the peer is
evaluated based on the clusters' relevance scores with respect to a
given query.

Although the clustering process is an offline process in the
framework, we still need an efficient clustering method that can
handle a large number of documents and adapt to updates on the
peers. In this paper, we adopt the widely used k-means clustering
algorithm \cite{Jain} to deal with these problems. It uses an
iterative procedure to find $K$ partitions of objects, which
minimize the total intra-cluster variance (or the squared error
function). Specifically, it starts with $K$ randomly selected
objects to serve as the centroids and divides the objects according
to the distances from them. Then it generates $K$ new centroids
based on the current partitions and starts another round of division
until a stable state is reached. Empirically, the k-means algorithm
can converge quickly and is considered to be very efficient. For
document space, k-means is to maximize the following measure:
\begin{equation} I = \sum_{i=1}^K {n_i \mu_i^T \mu_i} \end{equation}
where $n_i$ denotes the document number of the $ith$ cluster, and
$\mu_i$ the centroid of the $ith$ cluster.

It is proven that the solution to the k-means clustering method
coincides with the principal points solution \cite{Flury}, which
means it is a point-representation scheme where the best $K$
representative points (i.e. topics) are obtained. On the contrary,
SVD provides a component-representation of the document space and
ensures the best representation of the information content in a
reduced dimensional vector space. We will show that extending LSI
with clustering can especially adapt to FTR.

\section{FTR with C-DLSI}
In this section, we present the details about our cluster-based
distributed latent semantic indexing (C-DLSI) method for FTR. By
identifying the special properties of FTR, we extend LSI accordingly
to treat the peer selection issue, which overcomes the deficiencies
of the conventional approaches. In Section 4.1, we first formulate
the peer selection problem in FTR, analyze the properties, and
identify the deficiencies of the traditional approaches. Then we
propose the C-DLSI method especially tailored for FTR in Section
4.2. Based on C-DLSI, the corresponding descriptors and a federated
query processing in FTR are presented in Section 4.3 and Section
4.4, respectively. Finally in Section 4.5, we describe an update
scheme for a peer in a highly dynamic environment.

\subsection{Properties of Peer Selection in FTR}
In FTR, if the distribution of relevant documents across the results
returned by the peers were known, the peers could be ranked by the
number of relevant documents they return, which is known as
relevance-based ranking (RBR) \cite{Callan:0}. Consider a set of
peers $M=\{p_i\}$. To simplify the explanation, we assume that each
peer is required to return the top $N$ results, i.e., $\theta_i =
\{d_{ij}\}$ $(1 \leq j \leq N)$ with decreasing relevance score for
peer $p_i$. Let $r(d, q)$ denote the probability of relevance for
document $d$ given query $q$. The ranking value $r(p_i, q)$ for peer
$p_i$ can be represented as,
\begin{equation} r(p_i, q) = \sum_{j=1}^N r(d_{ij}, q) \end{equation}
Assume that the global weight of a term in each document is given.
There are two major issues here. One is how to determine a proper
relevance value $s(d_{ij}, q)$ which can approximate $r(d_{ij}, q)$
well. The other is how to summarize the descriptors of the peers
based on which the ranking value $r(p_i, q)$ can be properly
estimated.

A simple solution is to estimate the relevance score $s(d_{ij}, q)$
by computing the inner products between $d_{ij}$ and $q$ which is
called gGloss(0) \cite{Gravano:1} and can be described as,
\begin{equation} s(d_{ij}, q)=d_{ij}^T q \end{equation}
By only maintaining the document number $n_i$ and the centroid
$\mu_i$ of the peer $p_i$, the broker can estimate the ranking value
of peer $p_i$ as,
\begin{equation} r(p_i, q)= n_i \mu_i^T q \end{equation}
which is equal to $\sum_{i=1}^{n_i} s(d_{ij}, q)$. To solve the
problem of various $idf$ values of a term among peers, CORI
\cite{Callan}, on the other hand, estimates the ranking value by
using the term frequency of each query term in each peer. A further
improvement of CORI is to combine it with LSI \cite{Sogrin}. Since
these methods do not consider the topic space of the peers, the
effect of polysemy, i.e., some terms common to two conceptually
independent topics, is ignored. Different from document ranking,
peer ranking is suffered from the polysemy issue more seriously,
because the accumulation of small semantic deviations of the
documents may lead to big error in peer ranking. For example,
consider a collection of two documents. One is related to "apple,
fruit" while the other is related to "computation, math". If we
ignore polysemy, we may draw the conclusion that the set is related
to "apple computer", even though the individual relevance scores of
the documents with the query "apple computer" are not high.

A direct way to solve the polysemy issue is to use clustering to
identify the topics in a peer. Consider a conceptually homogenous
cluster. If it is regarded to be relevant to a query, say, "apple,
computer, product", then a document in it which does not contain any
query terms, e.g., talking about "MacBook, OS", is still likely to
be relevant to the query. Therefore, we should also consider the
synonyms in a cluster. Unfortunately, to the best of our knowledge,
none of the existing methods can adapt well to this situation. For
example, if we directly apply LSI on the whole collection and then
cluster the documents, then the polysemy can not be effectively
identified by LSI. \cite{Shen} tried to solve this problem by
representing a document with the descriptor of its cluster.
Specifically, the weight $\overline{w_t}$ of a term $t$ in the
descriptor of a cluster $c_l$ is computed by,
\begin{equation}\overline{w}_t = \frac{\sum_{d_{ij}\in c_l}{w(d_{ij}, t)}}{n_{l,t}} \end{equation}
where $w(d_{ij}, t)$ denotes the weight of term $t$ in document
$d_{ij}$ and $n_{i,t}$ the number of the documents in $c_i$ which
contain term $t$. We can see that, to handle synonyms, the weight of
a term in the descriptor is estimated only according to the
documents which contain them. Then similar formulas as (6) and (7)
are utilized to compute the ranking value of the peers. Similarly,
\cite{Xu} employs language model to determine the relevant cluster
and all of the documents in a relevant cluster are regarded to be
relevant. However, these methods are restricted by two major
drawbacks. First, they rely heavily on the quality of clustering and
do not consider the relations among clusters. Second, since they
assume that all the documents in a cluster is equally relevant, it
may exaggerate the relevance score of weakly relevant or irrelevant
document and is difficult to decide a proper ranking list of the
documents, which is known as a compatibility issue. To overcome
these limitations and adapt to the properties of FTR, we extend LSI
with clustering which considers the relations among clusters and
captures more accurate descriptions of the peers.

\subsection{Cluster-based Distributed LSI (C-DLSI)}
In our method, the collection of a peer is partitioned into a number
of clusters $\{c_i\}$ (using k-means clustering). Then, LSI is
employed to derive the semantic structure within each cluster, i.e.,
LSI space $C'_i = U'_i \Sigma'_i {V'_i}^T$ for cluster $c_i$ with
term-document matrix $C_i = U_i\Sigma_i V_i^T$. To make the LSI
spaces among clusters comparable, we restrict $C'_i$ with singular
values larger than a threshold $\varepsilon$. Let $\sigma_{i,j}$
denote the $jth$ singular value in $\Sigma_i$ and a number $k$
satisfy,
\begin{displaymath} \sigma_{i,j}\geq \varepsilon,~1<j\leq k \end{displaymath}
\begin{displaymath} \sigma_{i,j}<\varepsilon,~k<j\leq rank(C_i) \end{displaymath}
Then, the LSI space of $C_i$ is redefined as,
\begin{equation} C'_i = (C_i)_{\varepsilon} = (U_i)_k (\Sigma_i)_k {(V_i)_k}^T\end{equation}
Consider a diagonal block matrix $A$ for a peer with the form,
                      \begin{equation}A=\left(
                        \begin{array}{cccc}
                          C_1 &  &  &  \\
                           & C_2 &  &  \\
                           &  & \ddots &  \\
                           &  &  & C_K \\
                        \end{array}
                      \right)\end{equation}
where $C_i$ represents a conceptually independent topic, e.g.,
cluster $c_i$. It is easy to prove the following relation \cite{Li},
                      \begin{equation}{A_\varepsilon}=\left(
                        \begin{array}{cccc}
                          (C_1)_\varepsilon &  &  \\
                           &  \ddots &  \\
                           &  &  (C_K)_\varepsilon \\
                        \end{array}
                      \right)\end{equation}
It means if a collection is perfectly divided into a number of
conceptually independent topics and no polysemy exists, the LSI
space of a peer built in our method is equal to the traditional LSI
which is directly applied on the whole collection. Obviously, the
LSI spaces of the clusters can distinguish and capture the semantics
of the documents more precisely than the existing methods. In the
rest of this subsection, this idea will be further improved.

Each peer maintains the semantic structures of its clusters for
descriptor generation and federated query processing. Thus, we call
our method {\em cluster-based distributed LSI} (C-DLSI). Generally,
the clusters of a peer may have some relations from each other
because of several reasons, e.g., some topics are not conceptually
independent in nature or the clustering is not perfect enough and
some documents belonging to one topic are separated. In C-DLSI, a
semantic similarity measure between any two clusters (named paired
similarity) is introduced. This measure is estimated based on the
similarity of the LSI vector spaces and consequently, a network of
clusters is formed from words shared by each pair of clusters. With
the similarity network of clusters, C-DLSI can further exploit the
synonyms without loss of the polysemy information. Therefore, it can
especially adapt to FTR.

Similar to \cite{Bassu}, we define two levels of the paired
similarity. Consider two clusters $c_i$ and $c_j$ ($i\neq j$). Let
$T_i$ denote the term set for a cluster $c_i$ and $T_{ij}$ the
common term set for $c_i$ and $c_j$. The first level of paired
similarity $S1$ only captures the frequency of occurrence of common
terms. If $c_i$ and $c_j$ have common terms, we say there is a
direct link between them. Define the \emph{proximity} of $c_i$ and
$c_j$ to be the minimal number of the intermediate clusters which
link $c_i$ and $c_j$. Let $l$ denote the proximity between $c_i$ and
$c_j$, then $S1(c_i, c_j)$ is defined as (we only consider the case
of $l\leq 1$),
\begin{equation} S1(c_i, c_j)= (1/S_{ij}^l + l)^{-1} \end{equation}
where,
\begin{displaymath} S_{ij}^0 = \frac{|T_{ij}|^2}{|T_i| |T_j|} \end{displaymath}
\begin{displaymath} S_{ij}^1 = \max_{m} \frac{|T_{im}|^2 |T_{mj}|^2}{|T_i| |T_m|^2 |T_j|} \end{displaymath}

Moreover, a further level of the paired similarity captures the
semantics of the common terms, denoted as $S2$. Let $B_i = U'_i
{U'_i}^T$ denote the term similarity matrix for cluster $c_i$ and
$B_i^{Q}$ the restriction of $B_i$ to the term set $Q$. Define the
correlation measure between two $n\times m$ matrices $X$ and $Y$ as,
\begin{equation} R(X,Y) = |\frac{1}{nm}\sum_{i=1}^n{\sum_{j=1}^m{(\frac{X_{ij} - \overline{X}}{F_X})(\frac{Y_{ij} - \overline{Y}}{F_Y})}}| \end{equation}
where,
\begin{displaymath} \overline{X}=\frac{1}{mn} \sum_{i=1}^n {\sum_{j=1}^m
{X_{ij}}},~ \overline{Y}=\frac{1}{mn} \sum_{i=1}^n {\sum_{j=1}^m
{Y_{ij}}}
\end{displaymath}
\begin{displaymath} F_X = \frac{1}{mn} \sum_{i=1}^n {\sum_{j=1}^m
{X_{ij}^2}},~ F_Y = \frac{1}{mn} \sum_{i=1}^n {\sum_{j=1}^m
{Y_{ij}^2}}
\end{displaymath}
Then, $S2(c_i, c_j)$ is defined as,
\begin{equation} S2(c_i, c_j)= (1/S_{ij}^l + l)^{-1} \end{equation}
where,
\begin{displaymath} S_{ij}^0 = R(B_i^{T_{ij}},B_j^{T_{ij}}) \end{displaymath}
\begin{displaymath} S_{ij}^1 = \max_{m}{R(B_i^{T_{im}},B_m^{T_{im}})R(B_m^{T_{mj}},B_j^{T_{mj}})} \end{displaymath}
Based on these definitions, the paired similarity between $c_i$ and
$c_j$ is defined as,
\begin{equation}S(c_i, c_j)=S1(c_i, c_j) S2(c_i, c_j) \end{equation}

\begin{figure}
\centering \epsfig{file=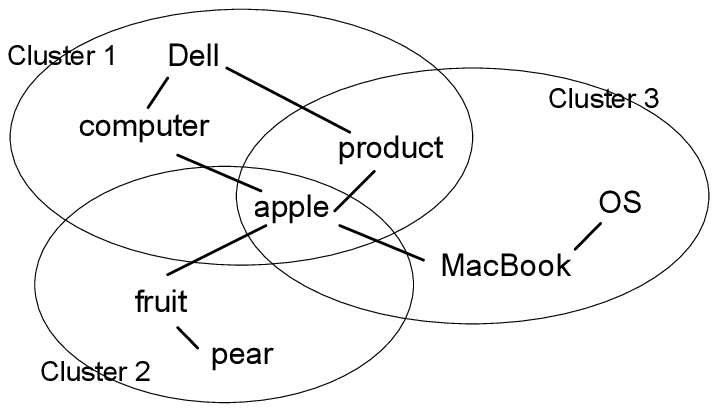, width=2.5in} \caption{An
example of the semantic structures in a collection.} \label{fig:ss}
\end{figure}

Before further explanation, we first consider an example of the
semantic structures in a collection as shown in Figure~\ref{fig:ss}.
Obviously, the polysemy of term "apple" can be identified via
clustering and divided into different clusters. However, some
conceptually relevant terms such as "MacBook" and "Computer" may
appear in different clusters. For a given query "computer", if only
the LSI space within a cluster is considered, then the relevant
documents in cluster 3 will be missed. For this reason, the paired
similarities among clusters are employed to extend the LSI space
built in a cluster.

For a cluster $c_i$, let $L_i = \{c_{i_1}, \ldots, c_{i_n}\}$ be a
set of clusters in the same peer as $c_i$ which satisfies the
following conditions,
\begin{displaymath} \forall c_{i_j}\in L_i,~S(c_i, c_{i_j}) > \delta, \end{displaymath}
\begin{displaymath} and~ \forall c_{i_{j1}},c_{i_{j2}}\in L_i,~j1<j2 \Rightarrow S(c_i, c_{i_{j1}}) \geq S(c_i, c_{i_{j2}}) \end{displaymath}
where $\delta$ is a threshold above which the cluster is regarded to
be relevant to $c_i$. We call $L_i$ the relevant clusters of $c_i$.
Given a query $q$ (denote $T_q$ as its term set), the relevance
score of cluster $c_i$ is computed as the sum of the relevance
scores of $c_i$ to all of the terms in $q$, which can be represented
as,
\begin{equation} s(c_i, q)= \sum_{t\in T_q}{s(c_i, t)} \end{equation}
The relevance score of cluster $c_i$ to a term $t$, i.e., $s(c_i,
t)$, can be estimated in two cases.

1) If $t \in T_i$, we have,
\begin{equation} s(c_i, t)= n_i \mu_{i,t} q_t \end{equation}
where $n_i$ is the number of documents in $c_i$ and $\mu_{i,t}$ is
the weight of $t$ in the centroid $\mu_i$ of the LSI space of $c_i$.

2) If $t \notin T_i$, the relevance score can not be derived
directly by using the LSI space of $c_i$. Then we can rely on its
relevant clusters $L_i$ to estimate the relevance score. Let
$c_{i_m}$ be the first cluster of $L_i$ which satisfies $t \in
T_{i_m}$. The projection of $t$ into the LSI space of $c_{i_m}$ can
be presented as,
\begin{equation}t' = B'_{i_m, t} q_t\end{equation}
where $B'_{i_m, t}$ denotes the column of $B'_{i_m}$ which
corresponds to term $t$. Therefore, the relevance score of $c_i$ to
term $t$ can be approximated as,
\begin{equation} s(c_i, t)= n_i \mu_i^T t'= n_i \mu_i^T B'_{i_m, t} q_t \end{equation}

With the relevance scores $\{s(c_i, q)\}$ of the clusters $\{c_i\}$
to the query $q$, we can estimate the ranking value of a peer $p$.
Let $c_1, \ldots, c_K$ be the clusters of $p$ with decreasing
relevance scores. Then the ranking value is estimated by considering
the most $h$ relevant clusters, which can be represented as,
\begin{equation} r(p, q)= \sum_{i=1}^h {s(c_i, q)} \end{equation}

Moreover, C-DLSI is efficient and scalable, since the size of a
cluster is substantially smaller than that of the entire collection.
With regard to document updates, it only requires reindexing of the
affected clusters. Thus, it is suitable for highly dynamic
environment in which documents are frequently updated.

\subsection{Descriptors of Peers}
A peer will publish a descriptor of its content to the broker for
peer selection. In C-DLSI, since a collection has been clustered,
the descriptor of a peer consists of a set of cluster descriptors.
According to Formulas $(16)\sim (20)$, we need at least the document
number $n_i$, the centroid of the LSI space $\mu_i$, and the eigen
matrix $U'_i$ to describe a cluster $c_i$. However, $U'_i$ is
usually quite large and may cause heavy communications between the
broker and the peers. To overcome this difficulty, we rewrite
Formula (19) as follows,
\begin{equation} s(c_i, t)= n_i \mu_i^T B'_{i_m, t} q_t = n_i (B'^T_{i_m, t} \mu_i)^T q_t\end{equation}
It means we only need a value $B'^T_{i_m, t} \mu_i$ instead of the
vector $B'_{i_m, t}$ to estimate the relevance score. For any term
$t$, a list of vectors $\rho_i = \{U'_{i_1} U'^T_{i_1}
\mu_i,\ldots,U'_{i_n} U'^T_{i_n} \mu_i)\}$, which correspond to the
order of the relevant clusters $L_i = \{c_{i_1}, \ldots, c_{i_n}\}$,
are guaranteed to find the value $B'^T_{i_m, t} \mu_i$. Therefore,
the transmission of the matrix $U'_i$ can be saved. Based on this,
we define the descriptor $D_i$ for cluster $c_i$ in C-DLSI to
contain the following aggregate information:

\begin{enumerate}
  \item The total number of documents in the cluster, $n_i$.
  \item The centroid of LSI space in the cluster, $\mu_i$.
  \item A list of vectors $\rho_i = \{U'_{i_1} U'^T_{i_1} \mu_i,\ldots,U'_{i_n} U'^T_{i_n}
  \mu_i)\}$, which correspond to the order of the relevant clusters $L_i = \{c_{i_1}, \ldots,
  c_{i_n}\}$.
\end{enumerate}

\noindent That is,
\begin{equation}D_{c_i} = \{ N_i, \mu_i, \rho_i \} \end{equation}
Then we define the descriptor $D_p$ for peer $p$ as,
\begin{equation}D_p = \{ D_{c_i} | c_i\in p \} \end{equation}
which represents the peer with more fine-grained descriptions of its
clusters. Since the number of clusters $K$ is extremely small,
publishing this descriptor causes little overhead in C-DLSI.

Note that we assume in this paper that there is little or no overlap
among the peers. This is a reasonable assumption in most cases.
Since each peer is supported to cover a different part of the web,
it corresponds to a distinct database. When this assumption is
violated, we can add more aggregate information into cluster
descriptors as proposed in \cite{Bender:1}.

\subsection{Federated Query Processing}
As described in Section 3, federated query processing in FTR
contains three steps, namely peer selection, local text retrieval,
and result merging. In this subsection, we will discuss these three
issues in C-DLSI.

As discussed in Section 4.2, the broker compute the ranking values
for all the peers according to Formula (20) based on the
descriptors. In particular, when computing the relevance score of a
cluster $c_i$ to a term $t$ of the query $q$ with $t \notin T_i$,
the broker will scan the list $\rho_i$ and find the first relevant
cluster $c_{i_m}$ which contains $t$. Then the relevance score
$s(c_i, t)$ is computed by,
\begin{equation} s(c_i, t)= n_i \rho^t_{i, m} q_t \end{equation}
where $\rho^t_{i, m}$ denotes the weight of $t$ in the $mth$ vector
of $\rho_i$. Otherwise, the relevance score $s(c_i, t)$ can be
directly computed according to Formula (17). The broker will choose
the peers with largest ranking values and forward the query $q$,
together with the IDs of the $h$ most relevant clusters, to each of
them. Then it enters the phase of local retrieval.

Local retrieval is performed by peers to retrieve relevant documents
from the collections. C-DLSI only considers the LSI spaces of the
$h$ most relevant clusters specified by the broker. The relevance
score of a document $d_j$ to query $q$ is evaluated based on its LSI
vector $d'_j$ in the corresponding cluster $c_i$. Similarly, we
have,
\begin{equation}s(d_j, q)= \sum_{t\in T_q}{s(d_j, t)} \end{equation}
If $t \in T_i$, then the relevance score of the document $d_j$ to
term $t$ can be computed by,
\begin{equation}s(d_j,t) = d'_{j,t} q_t \label{eqn:sim}\end{equation}
where $d'_{j,t}$ denotes the weight of $t$ in $d'_j$. Otherwise, the
first cluster in $L_i$ which contains term $t$ will be found,
denoted as $c_{i_m}$ and the relevance score is estimated as,
\begin{equation}s(d_j,t) = d'^T_j B'_{i_m, t} q_t \end{equation}
Thus, the evaluated documents can be sorted according to their
relevance scores, and only the top-ranked documents will be returned
as the results to the broker.

Result merging in FTR tries to provide a uniform ranked list of the
documents returned from multiple peers. Assume that each peer has
the global weights for all of the terms in the documents and applies
the same ranking function. Then the relevance score of a document
estimated by the peer is also valid as a global score among all
peers. Thus, we can simply merge the documents according to their
relevance scores returned by the corresponding peers in C-DLSI.

Another factor considered in our framework is the compatibility
between peer selection and local text retrieval. Since the goal of
FTR is to retrieve valuable peers that can return most relevant
documents, this process is also impacted by local text retrieval
schemes. Thus it requires the peer selection and local text
retrieval to be compatible and consistent, which is called the
compatibility issue. In C-DLSI, we try to guarantee this property in
peer selection method and local text retrieval. Previous research
has shown that LSI can help improve document retrieval in a single
collection. However, most conventional methods for peer ranking are
more likely to select the peers with the largest number of weighted
keywords, which does not conform to the basic principle of LSI.
Moreover, though several peer selection methods consider the
semantic structures of the collections, they ignore the
compatibility issue or it is hard to find a proper local text
retrieval method to adapt to their peer ranking scheme.

\subsection{Collection Update}
Although our C-DLSI method employs LSI in a distributed way and only
requires applying SVD in a relatively small scale (i.e., on clusters
only), collection update could still be costly for extremely large
and dynamic collections. In our framework, we utilize a lazy scheme
to handle this problem. In particular, each peer keeps all of the
semantic transformation matrices $\{U'_i\}$ of its own clusters
(refer to Section 4.2). When an update occurs, e.g., due to textual
update or newly crawled documents, it will first assign the updated
documents to the most related clusters, e.g., cluster $c_i$, and
then directly use $U'_i$ to evaluate the semantic vector $d'_j$
according to the following formula:
\begin{equation}d'_j = U'_i U'^T_i d_j\end{equation}
where $d_j$ is the updated document vector. It can be easily proven
that this form is consistent with the original form of Formula~(2).
If the amount of updates exceeds a threshold for a cluster, the
corresponding peer will rebuild its LSI by applying SVD on the
cluster again. This update handling scheme is also tested and
analyzed in the experiments.

\begin{table}
\centering \caption{Summary of the experimental data}
\begin{tabular}{|c|c|c|} \hline
Type & Count & Avg. Length\\ \hline Document & 53595 & 213.455\\
\hline Query & 50 & 2.740\\ \hline Term & 186319 & -\\
\hline\end{tabular}
\end{table}

\begin{figure}
\centering \epsfig{file=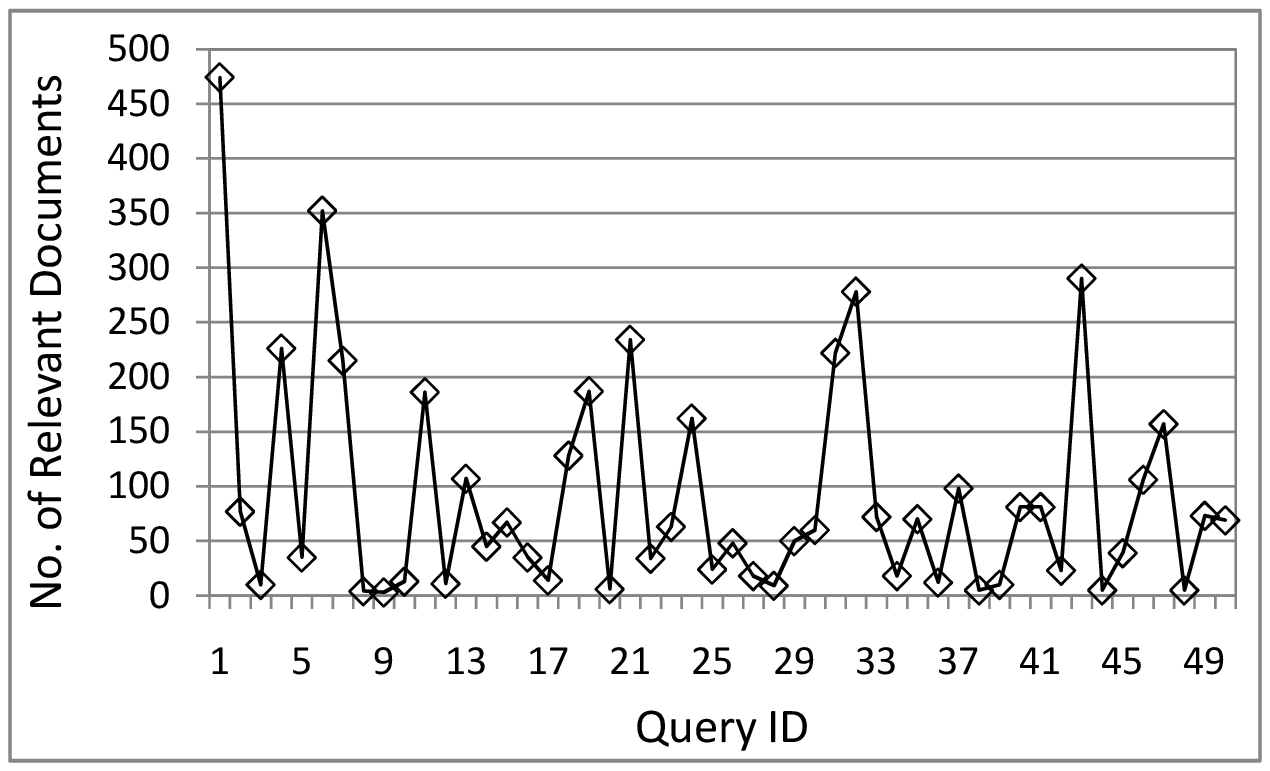, width=2.5in} \caption{Number
of relevant documents for each query.}
\end{figure}

\section{Experiments}
To evaluate the C-DLSI method, we build a simulation platform with
one broker and $50$ peers. The documents come from the TREC
collection Volume 4 and Volume 5, which consist of over $500,000$
documents with about $2.1GB$ in total size. In this section, we will
first present the setup of the experiments, and then show the
results from the C-DLSI evaluation.

\subsection{Experimental Setup}
In our experimental platform, we use the documents of the TREC
collection Volume $4$ and Volume $5$. The queries are extracted from
TREC-6 ad hoc topics (topics $301-350$). To simulate short Web
queries, we use the terms appearing in the Title field of the topic
description as the keywords. In the following experiments, we will
also discuss the effect of query length on C-DLSI. Moreover, the
standard relevance judgments provided by NIST are used to evaluate
the retrieval effectiveness. Since only a portion of the documents
are manually judged, we select them as the indexed documents to make
the evaluation more reasonable. In particular, the selected
documents are uniformly distributed to $50$ indexing collections,
which is considered the hardest scenario compared to a skewed
distribution \cite{Shen}. Table 1 gives a summary of the data set
used in the experiments. Figure 4 and Figure 5 show the statistics
of the indexed documents for each query. We can see that the number
of relevant documents for each query is relatively small and it is
not easy to identify them for most queries.

\begin{figure}
\centering \epsfig{file=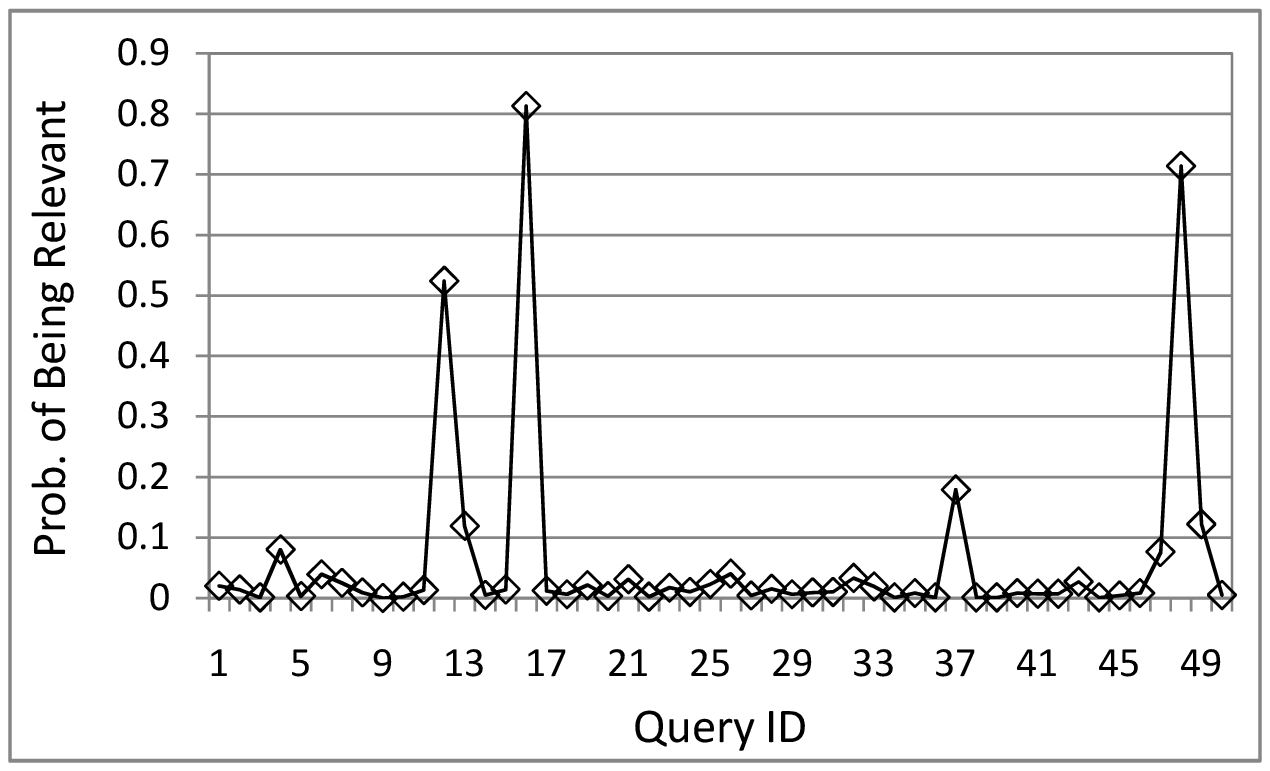, width=2.5in} \caption{The
probability of a document which contains keywords being relevant for
each query.}
\end{figure}

\begin{table}
\centering \caption{Parameters used in the experiments}
\begin{tabular}{|c|c|c|} \hline
Notation & Description & Values\\
\hline $K$ & Cluster $\#$ & $10, 20, 30$ \\
\hline $N$ & Retrieved doc $\#$ in each peer & $10, 20$ \\
\hline $G$ & Selected peer $\#$ (cast number) & $5, 10, \ldots, 50$ \\
\hline $\varepsilon$ & Threshold of LSI & $1, 1.5, \ldots, 9$ \\
\hline $h$ & Number of relevant clusters $\#$ & $5, 10, 15, 20$ \\
\hline
\end{tabular}
\end{table}

\begin{figure*}
\centering \epsfig{file=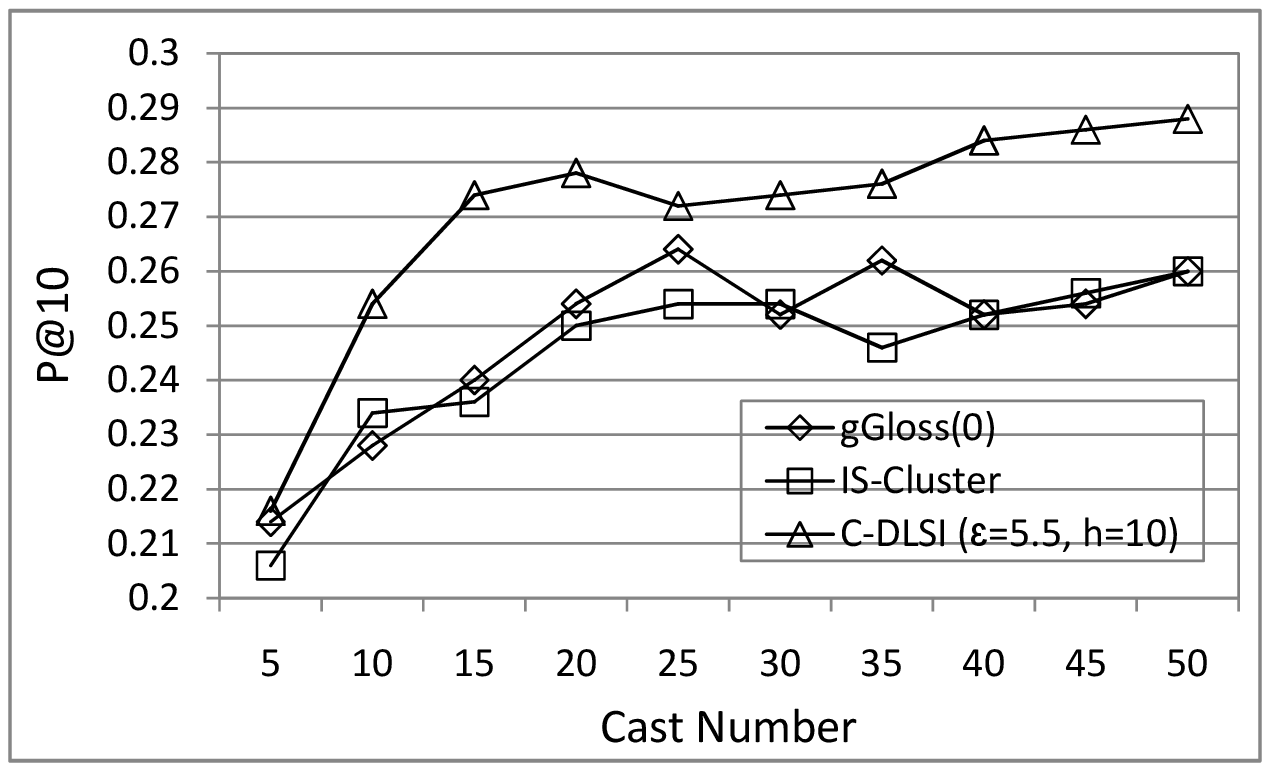, width=1.7in}
\epsfig{file=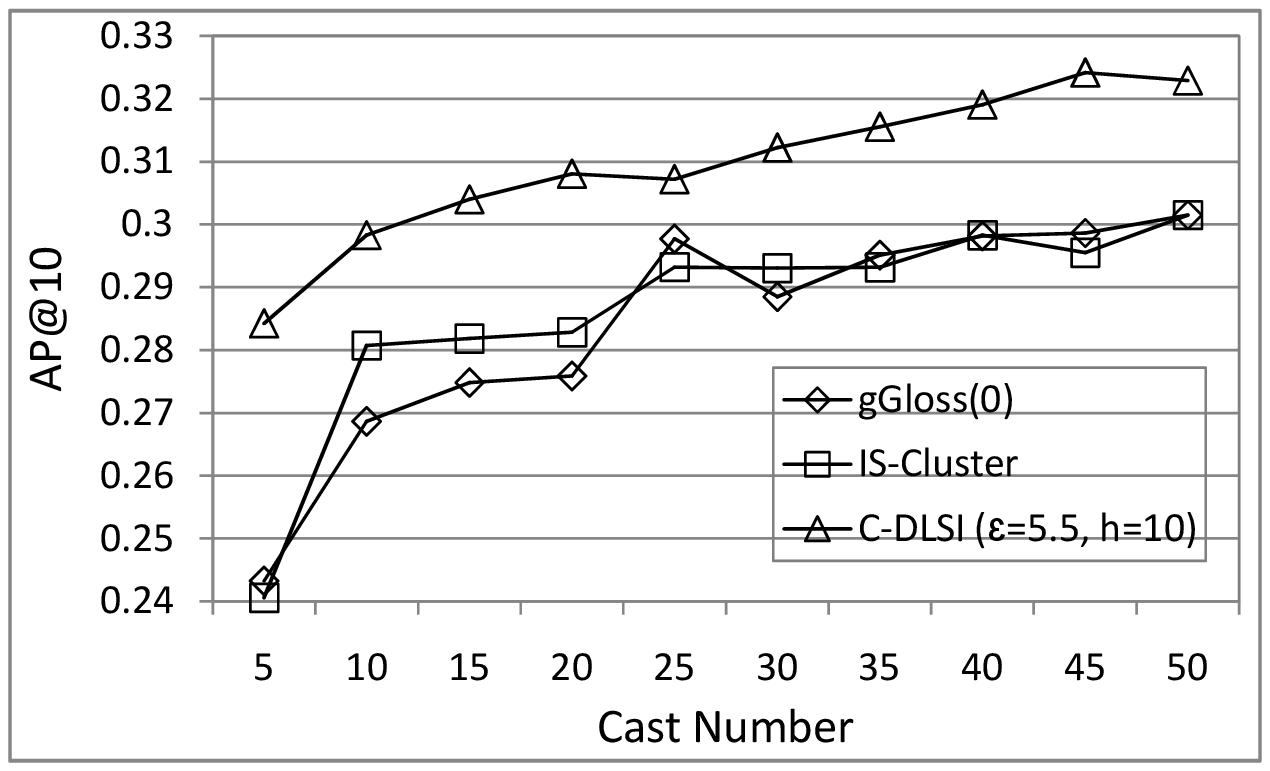, width=1.7in}
\epsfig{file=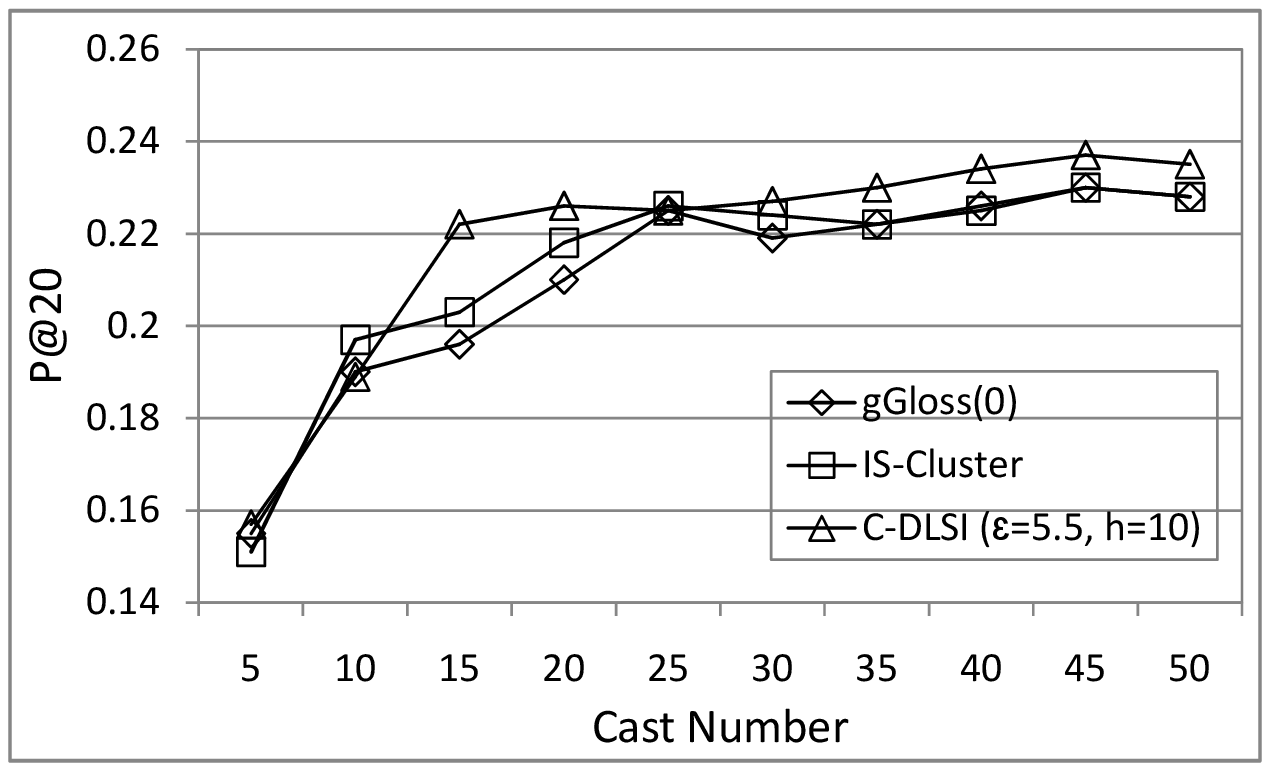, width=1.7in}
\epsfig{file=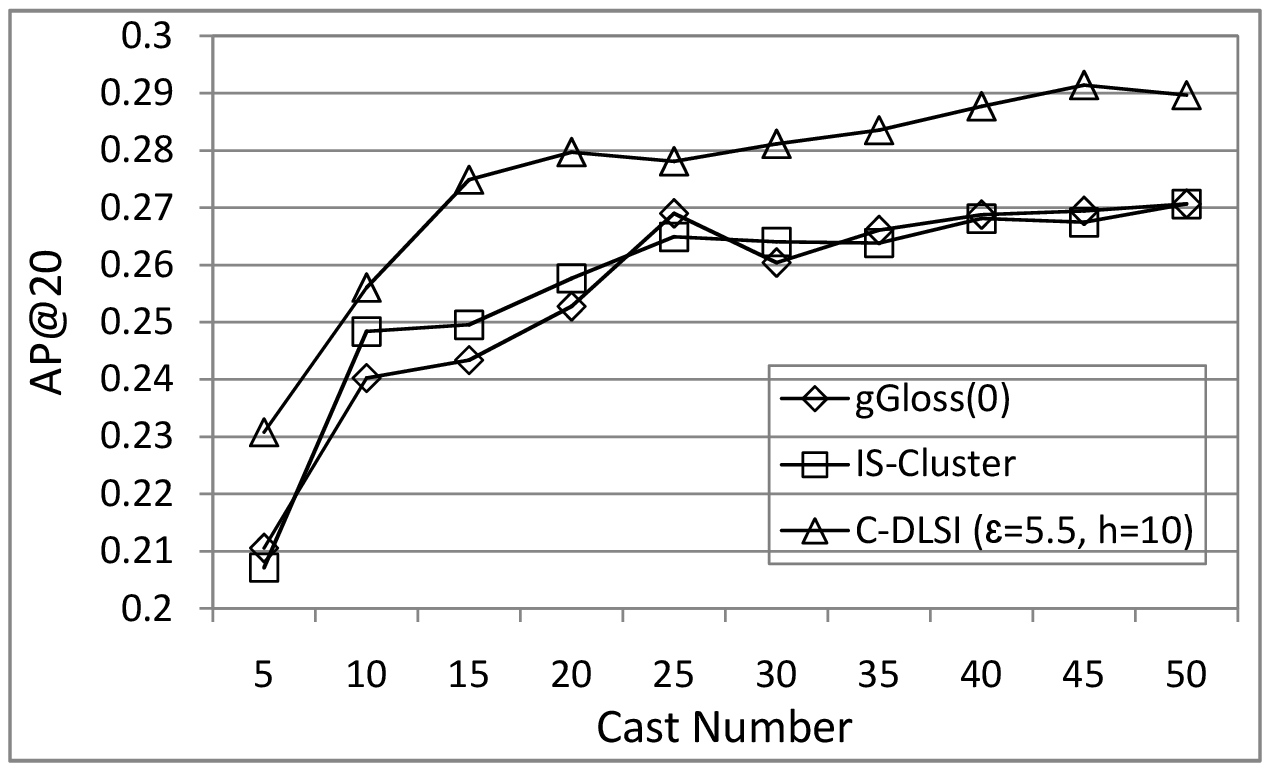, width=1.7in} \caption{Performance
comparison w. r. t. increasing cast number for $K = 10$.}
\end{figure*}

\begin{figure*}
\centering \epsfig{file=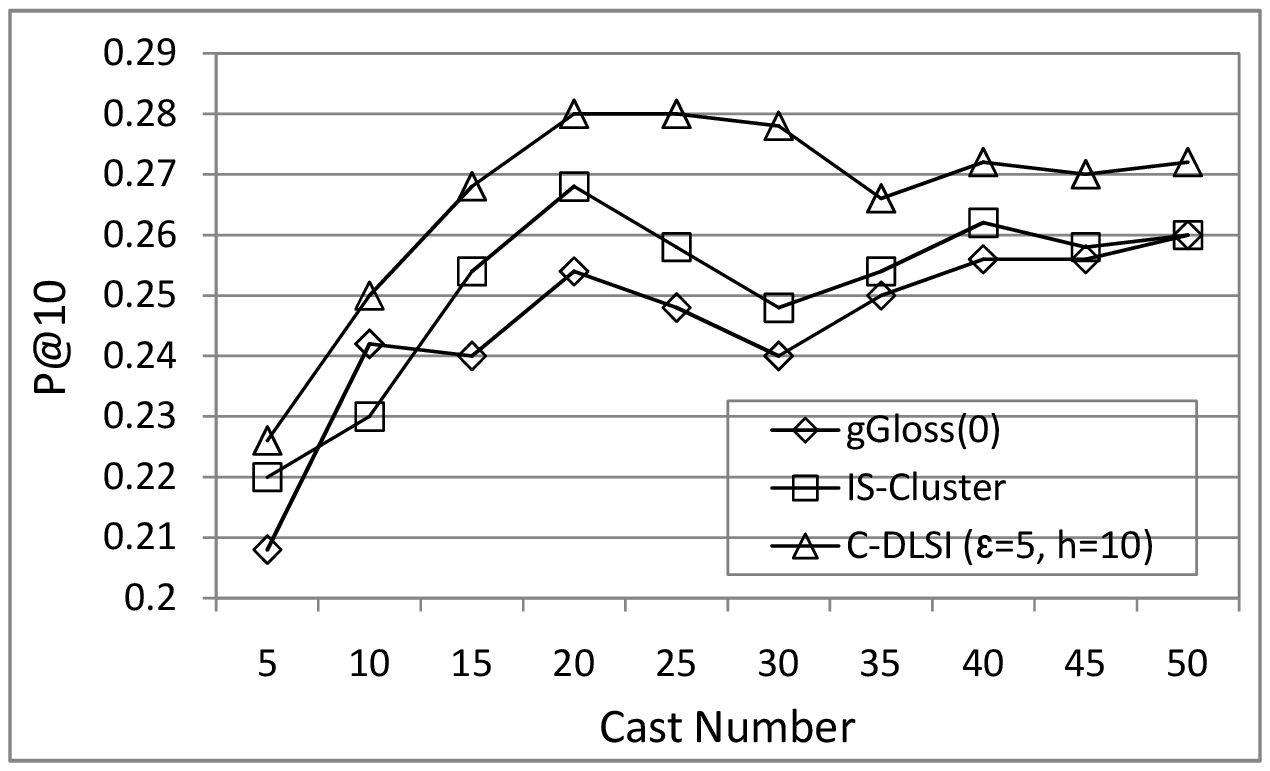, width=1.7in}
\epsfig{file=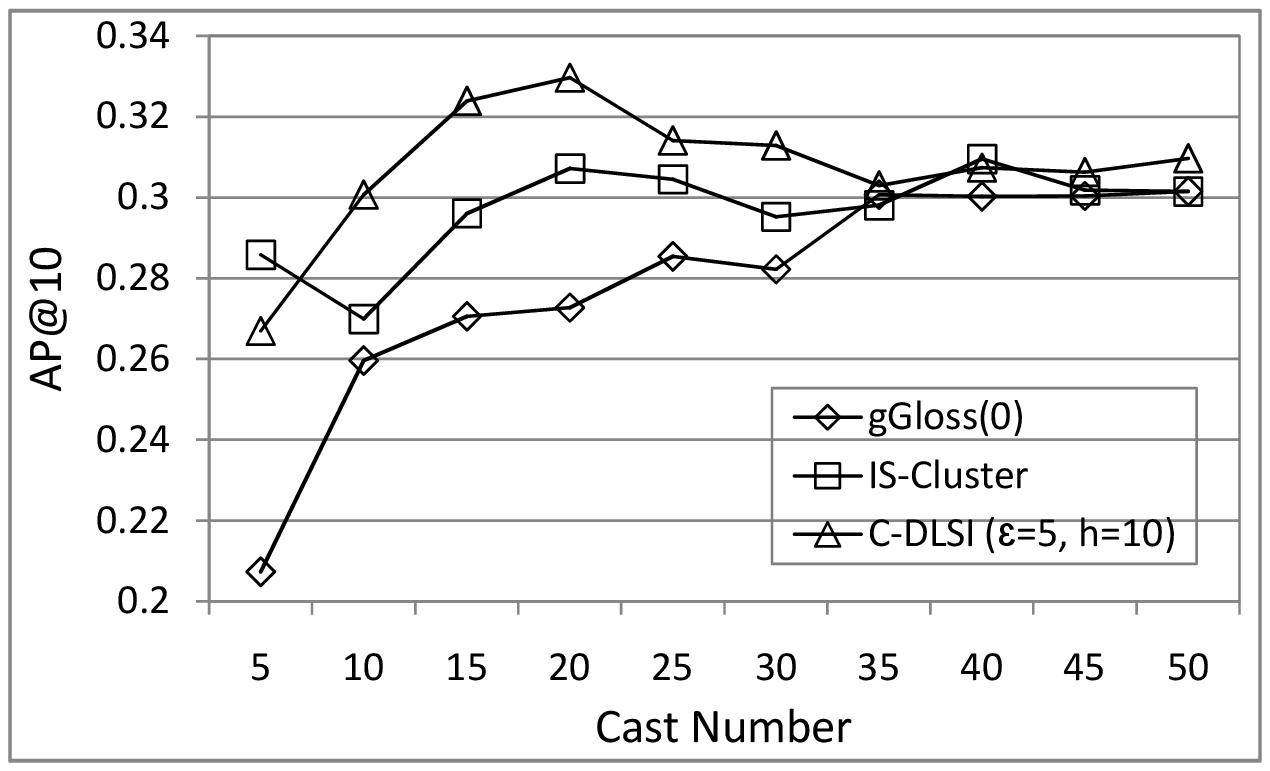, width=1.7in}
\epsfig{file=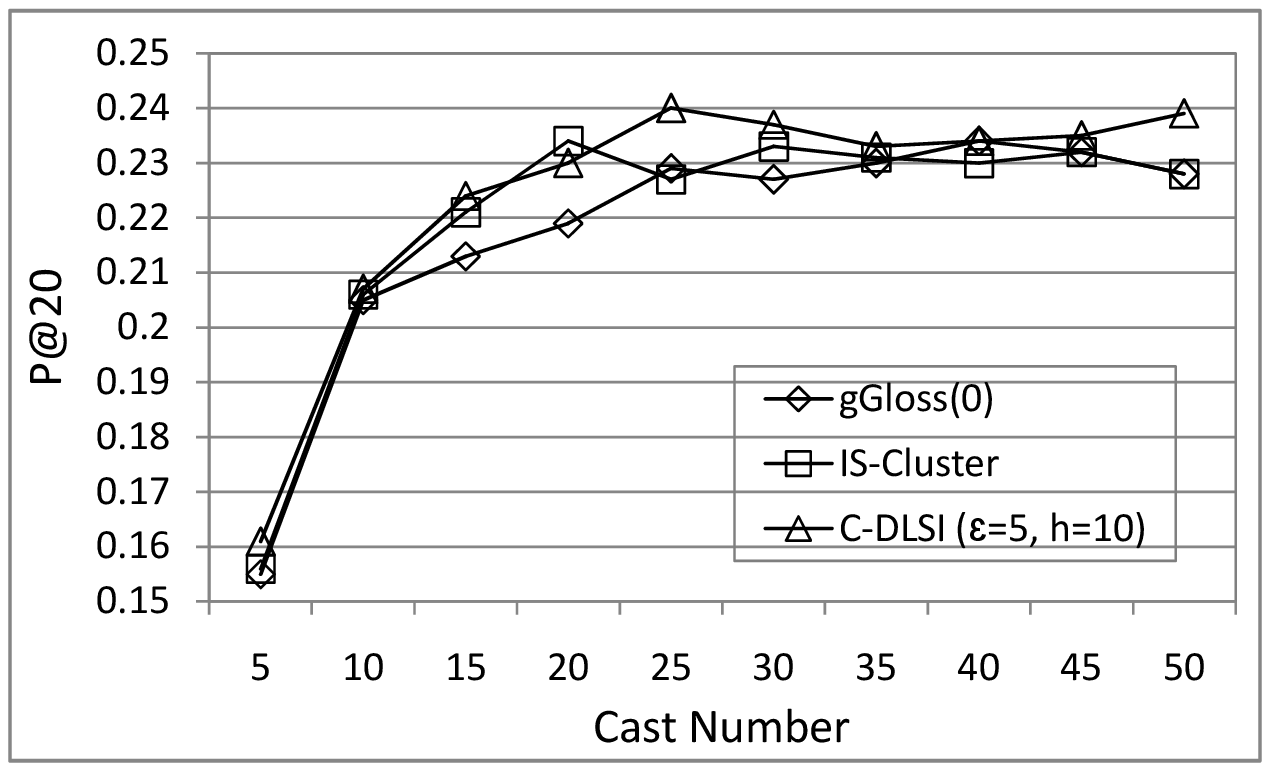, width=1.7in}
\epsfig{file=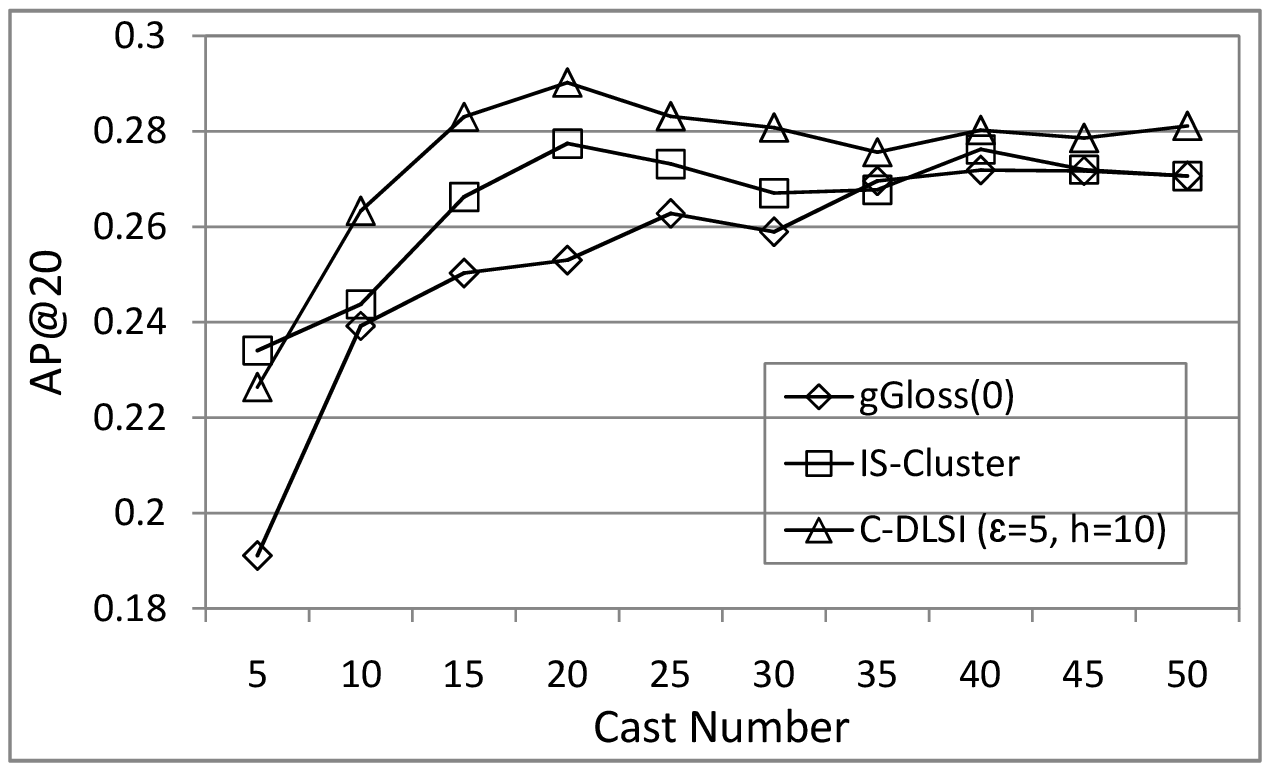, width=1.7in} \caption{Performance
comparison w. r. t. increasing cast number for $K = 20$.}
\end{figure*}

We use the LogEntropy weighting scheme \cite{Dumais} to compute the
weight vector of each document, which is defined as
\begin{displaymath}w_{ij} = [\log_2{(1+tf(i, j))}]\cdot [1+\sum_j{\frac{p_{ij}\log_2{p_{ij}}}{\log_2n}}] \end{displaymath}
\noindent where $tf(i,j)$ is the frequency of term $i$ in document
$j$, $n$ is the total number of documents in the collection, and
$p_{ij}=tf(i,j)/\sum_j{tf(i,j)}$. The parameters and their settings
used in the experiments are shown in Table 2. Generally, the
effectiveness of a FTR system is not evaluated by the precision at
recall points. Since only a subset of the peers is selected, it is
usually impossible to retrieve all of the relevant documents. As in
other research works \cite{Si:1}, we use two metrics to evaluate the
quality of the merged results. One is the top-N precision (P@N),
which can be defined as follows.
\begin{equation}P(q,N)=\frac{|R(N)|}{N}  \end{equation}
\noindent where $R_N$ stands for the set of relevant documents in
the top $N$ results. As a complement, we also use another metric
named top-N average precision (AP@N) to evaluate the distribution of
the relevant documents in the top-N results. AP@N is defined as,
\begin{equation}AP(q,N)=\frac{\sum_{i=1}^N{P(q,i)}}{N}\end{equation}
which indicates that the higher the relevant documents are ranked,
the larger AP@N will be. Unless stated to the contrary, the
evaluation metrics shown in this paper are the average for all $50$
queries. For comparison, as mentioned in Section 2, we also
implemented two baseline algorithms gGloss(0) \cite{Gravano:1} and
IS-Cluster \cite{Shen} that were shown to be very effective.

\subsection{Experimental Results}
In the experiments, we first evaluate the performance of C-DLSI and
study the impacts of each parameter. Then we analyze the
compatibility issue for C-DLSI in FTR. Next, we compare our method,
denoted as C-DLSI($\varepsilon$), with another form of C-DLSI which
is based on a truncated value $k$, namely, C-DLSI($k$). Finally, we
examine the performance of the collection update algorithm. Since
FTR usually selects a small number of peers, we focus more on the
performance for small cast numbers (e.g., $T\leq 25$) in the
experiments.

\begin{table}
\centering \caption{Performance comparison for the three methods on
peer selection}
\begin{tabular}{|c|c|c|c|c||c|c|c|} \hline
\multirow{2}{*}{T} & \multicolumn{2}{|c|}{Comp0} &
\multicolumn{2}{|c||}{Comp1} & \multicolumn{3}{|c|}{Avg. Recall} \\
 & $>$ & $<$ & $>$ & $<$ & gGloss(0) & IS-Cluster & C-DLSI \\
\hline 5 & \textbf{0.56} & 0.22 & \textbf{0.44} & 0.24 & 0.149 & \textbf{0.165} & 0.159  \\
\hline 10 & 0.4 & \textbf{0.42} & \textbf{0.4} & 0.32 & \textbf{0.288} & 0.282 & 0.28 \\
\hline 15 & \textbf{0.46} & 0.36 & 0.42 & 0.42 & 0.398 & 0.403 & \textbf{0.412} \\
\hline 20 & \textbf{0.38} & 0.36 & 0.38 & \textbf{0.5} & \textbf{0.521} & 0.52 & 0.51 \\
\hline 25 & 0.32 & \textbf{0.46} & 0.36 & \textbf{0.5} & 0.608 & \textbf{0.624} & 0.604 \\
\hline 30 & 0.34 & \textbf{0.46} & 0.36 & \textbf{0.4} & 0.686 & \textbf{0.702} & 0.698 \\
\hline 35 & 0.3 & \textbf{0.44} & 0.4 & \textbf{0.42} & 0.776 & \textbf{0.782} & 0.768 \\
\hline 40 & 0.34 & \textbf{0.4} & \textbf{0.42} & 0.38 & 0.853 & \textbf{0.859} & 0.858 \\
\hline 45 & 0.28 & \textbf{0.34} & \textbf{0.36} & 0.3 & \textbf{0.938} & 0.932 & 0.932 \\
\hline 50 & 0.0 & 0.0 & 0.0 & 0.0 & 1.0 & 1.0 & 1.0 \\
\hline
\end{tabular}
\end{table}

\subsubsection{Performance Evaluation}
At the beginning, the descriptors of all $50$ peers are stored in
the broker. During the experiments, the broker will load the short
queries extracted from TREC topics $301-350$ and perform peer
selection and final result merging. Table 3 presents the peer
selection results for each approach, where the setting of C-DLSI is
$K=20$, $\varepsilon=5$, and $h=10$. The comparison criterion is
based on the number of relevant documents contained in the selected
peers. Comp0 and Comp1 compare C-DLSI with gGloss(0) and IS-Cluster
respectively, by measuring the portion of the queries in which one
method outperforms the other. Specifically, $>$ means C-DLSI
outperforms gGloss(0) (IS-Cluster), while $<$ denotes the reverse.
Avg. Recall denotes the average recall of the selected peers for all
of the $50$ queries. We can see that C-DLSI as a whole outperforms
gGloss(0) and is close to IS-Cluster. Similar results can be
observed for other settings.

\begin{figure}
\centering \epsfig{file=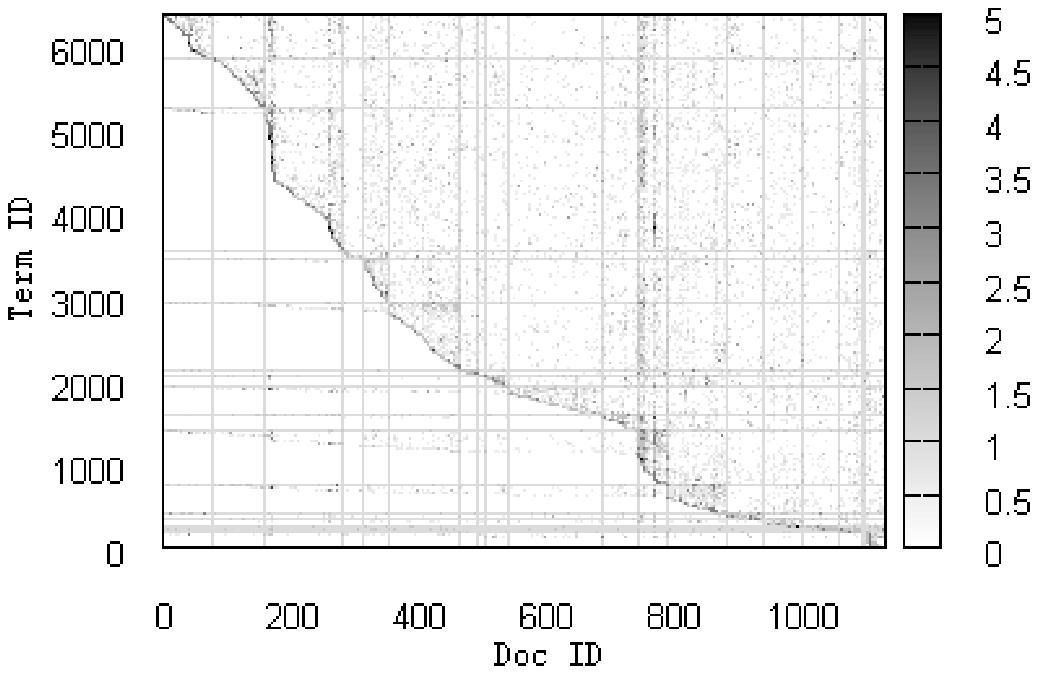, width=2.5in} \caption{An
example of gray-scale map for a peer when $K = 20$. The points
represent the weights of the terms. The vertical lines separate the
clusters.}
\end{figure}

For performance comparison, Figure 5 and Figure 6 show the top-N
precision (P@N) and top-N average precision (AP@N) of all three
methods for increasing cast number under two different settings. We
can see that C-DLSI with a proper threshold $\varepsilon$ (discussed
in Section 5.2.2), e.g., $\varepsilon = 5.5$ for cluster number $K =
10$, in general outperforms the other two methods under both
evaluation metrics. To understand these results further, we check
the characteristics of the peers in the simulation. Specifically,
the gray-scale map of a peer for $K=20$ is given in Figure 7. In
this map, some popular terms are removed. It shows that the TREC
data used in the experiments are relatively sparse, which means it
is generally difficult to properly rank them. Besides, it leads to
unsatisfactory clustering results. Based on this, we can only expect
a modest improvement by applying C-DLSI on this dataset. Note that
we use different collection assignments in two experiments to gain a
general conclusion. The determination of parameters are discussed in
Section 5.2.2. Since the performance of C-DLSI in different
collection assignments in general are similar, we mainly utilize the
collection assignment of $K=20$ as an example to investigate the
properties of C-DLSI in the following experiments.

\begin{figure}
\centering \epsfig{file=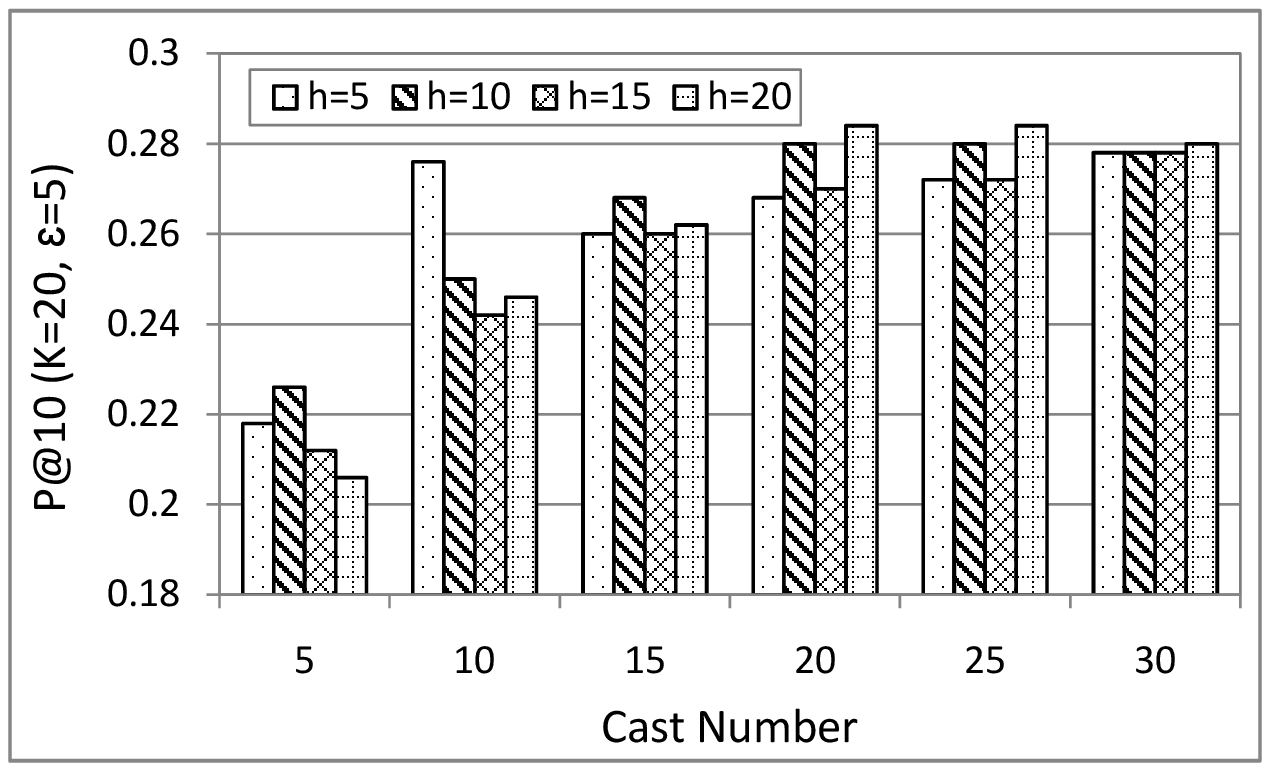, width=2.5in} \caption{C-DLSI
performance (P@10) with different $h$ values for increasing cast
number. }
\end{figure}

\subsubsection{Impacts of Parameters}
\subsubsection*{Number of relevant clusters $h$}
In C-DLSI, only $top-h$ relevant clusters are used to judge the
relevance of a collection (see Section 4.5). Figure 8 shows the
performance of C-DLSI with different $h$ values for cast number 5,
10, 15, 20, 25 and 30. From the results, we can see that for the
smallest number $T=5$, the smallest $h$ value (i.e., $h=5$)
outperforms other settings. As the cast number increase, larger $h$
values gradually become more preferable. It supports the fact that
clustering is able to identify the documents which contain query
terms but are irrelevant to the query, e.g., by keeping them in the
clusters that refer to other topics. Thus, with smaller $h$, we can
remove the impacts of these irrelevant documents and provides better
relevance estimation of the peers. That is why this strategy can
achieve higher precision for most relevant peers (i.e., small cast
numbers). However, because of the limited clustering quality on the
peers which are not quite relevant to the query, some relevant
documents may be assigned to the wrong clusters, that refer to other
topics. In this case, exploring more clusters, which is achieved by
having larger $h$ values, will be more effective. Therefore, larger
$h$ gains better performance as more peers are considered (i.e.,
larger cast numbers). In the experiments, we choose $h=10$ for the
case of $K=20$ to analyze the performance of C-DLSI.

\begin{figure}
\centering \epsfig{file=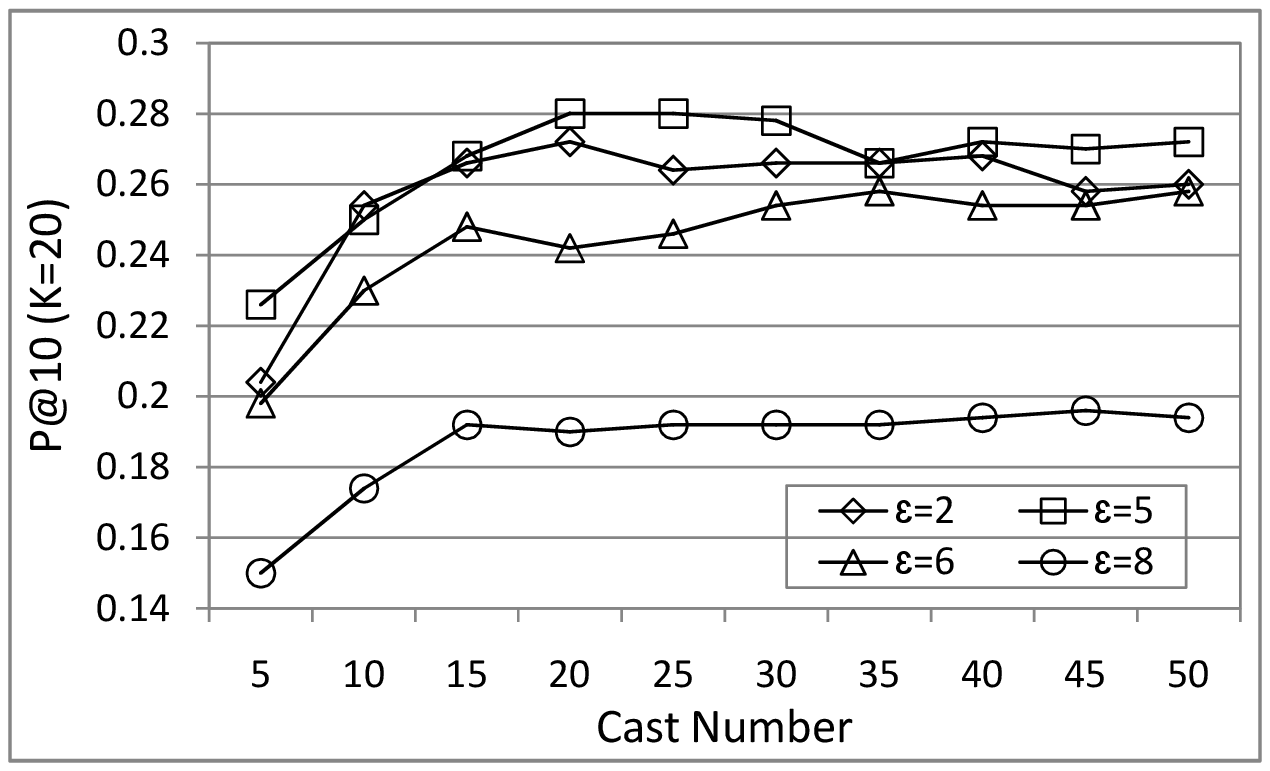, width=2.5in} \caption{C-DLSI
performance (P@10) with different $\varepsilon$ values for
increasing cast number. }
\end{figure}

\begin{figure}
\centering \epsfig{file=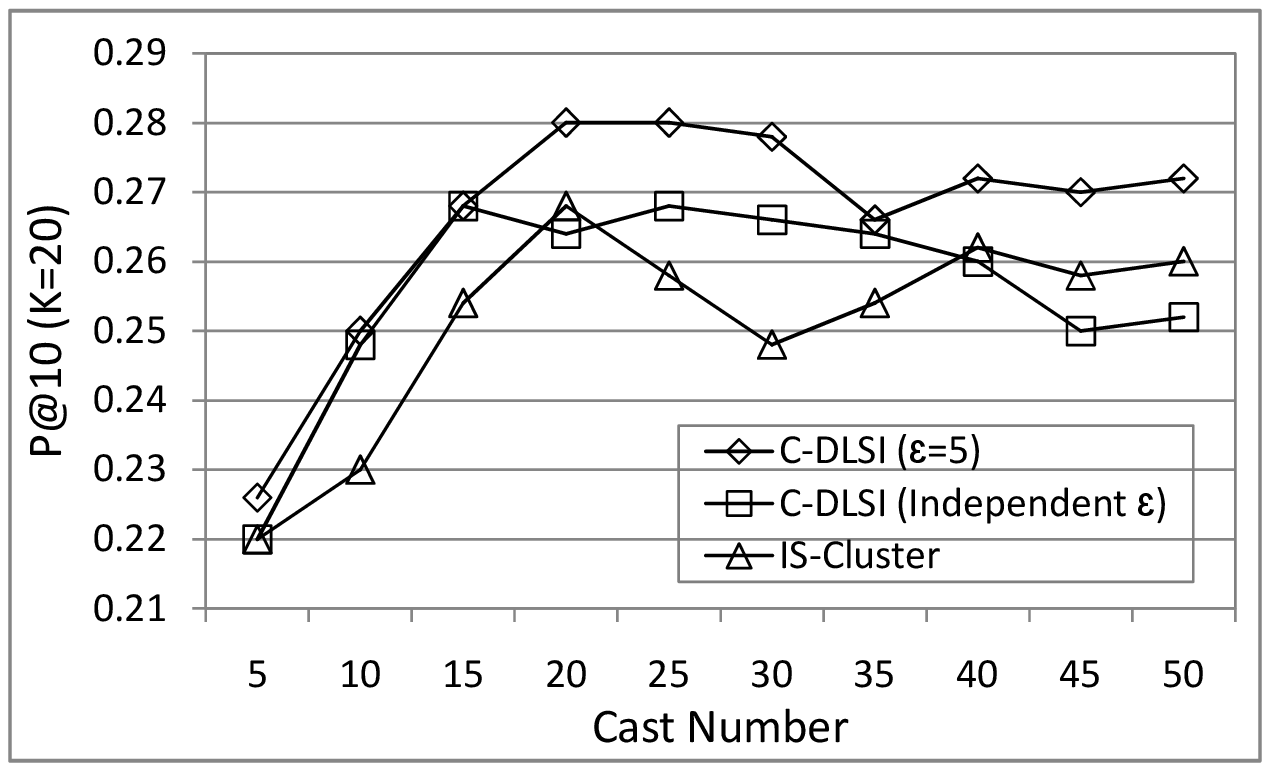, width=2.5in} \caption{C-DLSI
performance for the case when peers select their $\varepsilon$
values independently. The performances of other methods such as
C-DLSI($\varepsilon=5$) and IS-Cluster are also given for
comparison. }
\end{figure}

\subsubsection*{Threshold $\varepsilon$ of LSI}
Basically, if the threshold $\varepsilon$ of LSI is zero, then
C-DLSI will degenerate to a pure cluster-based peer selection
method. In the experiments, we can see from Figure 9 that the
effectiveness of C-DLSI is not monotonously increasing with
$\varepsilon$. Further, it always reaches the highest in the middle.
In the example, the optimal value of $\varepsilon$ is $5$ for
$K=20$. That is why C-DLSI with proper threshold $\varepsilon$ can
improve the performance of the federated querying processing in FTR.
Another important issue here is how to decide the proper value of
$\varepsilon$ for each peer. Since each search peer in FTR maintains
their own LSI, the threshold $\varepsilon$ has to be decided
independently. In Figure 10, we consider this issue in the same
situation as in Figure 9 and make each peer select their local
optimal $\varepsilon$ independently. Based on the metric of AP@10,
each peer searches the optimal threshold $\varepsilon$ from the
testing interval between 1 and 9, which finally concentrates on
either 1 or 5. From the results, we can see that this threshold
decision method causes a slight performance decrease compared to the
optimal case of $\varepsilon=5$ but still outperforms the other
methods such as IS-Cluster. Generally, how to automatically decide
the proper $\varepsilon$ value for each peer and achieve a global
optimal performance remains a open question to be answered in our
future work.

\begin{figure}
\centering \epsfig{file=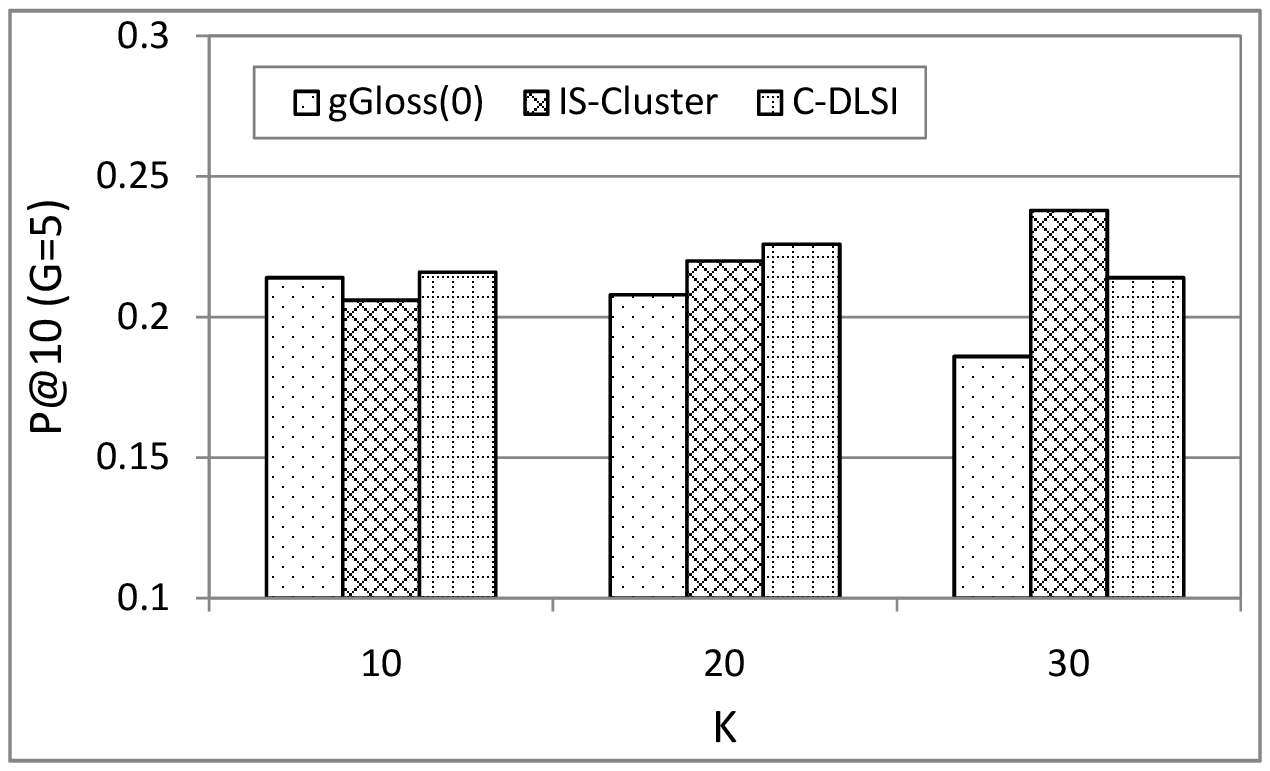, width=1.7in}
\epsfig{file=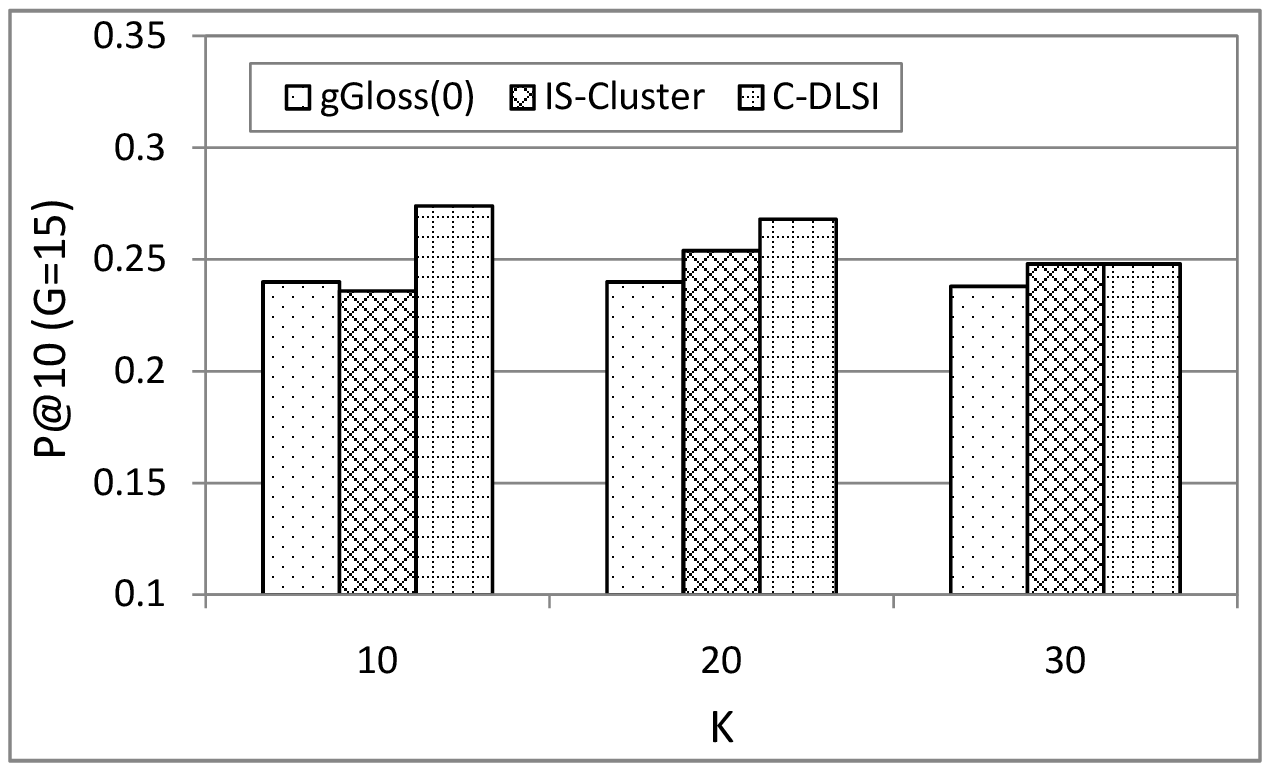, width=1.7in} \epsfig{file=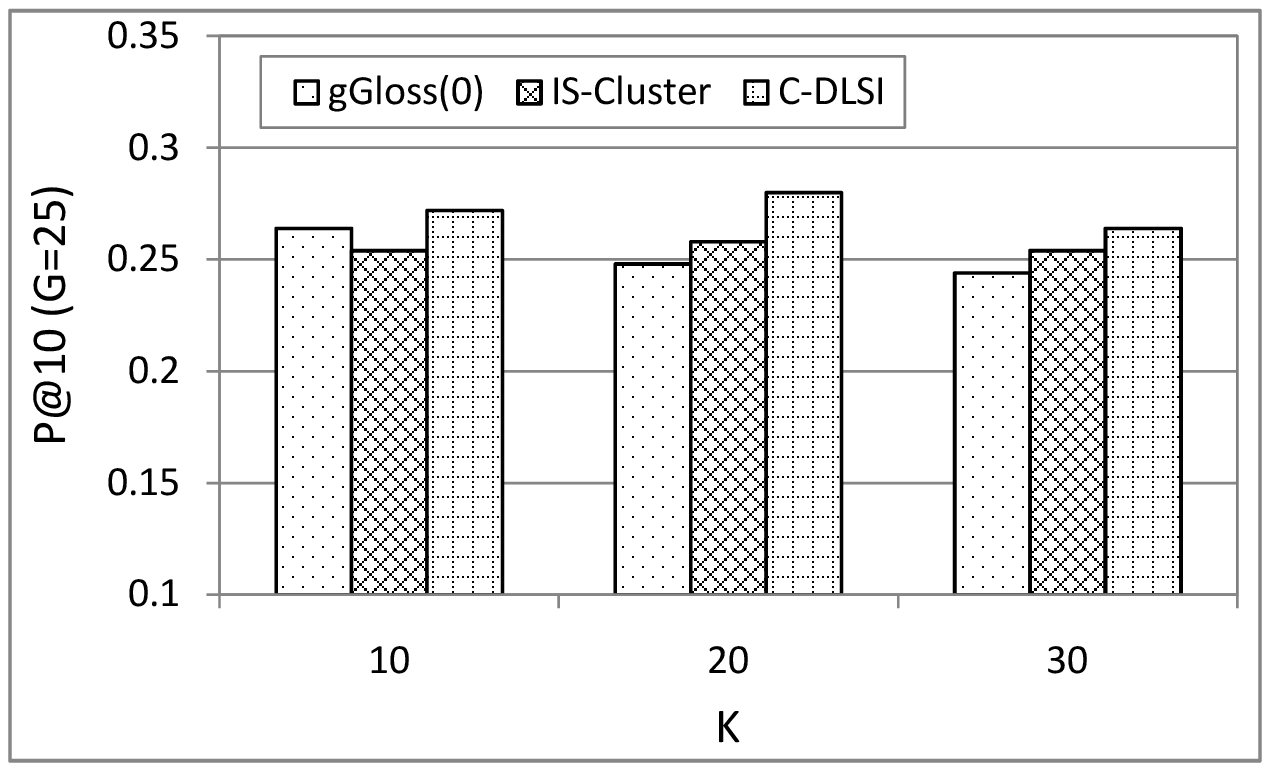,
width=1.7in} \epsfig{file=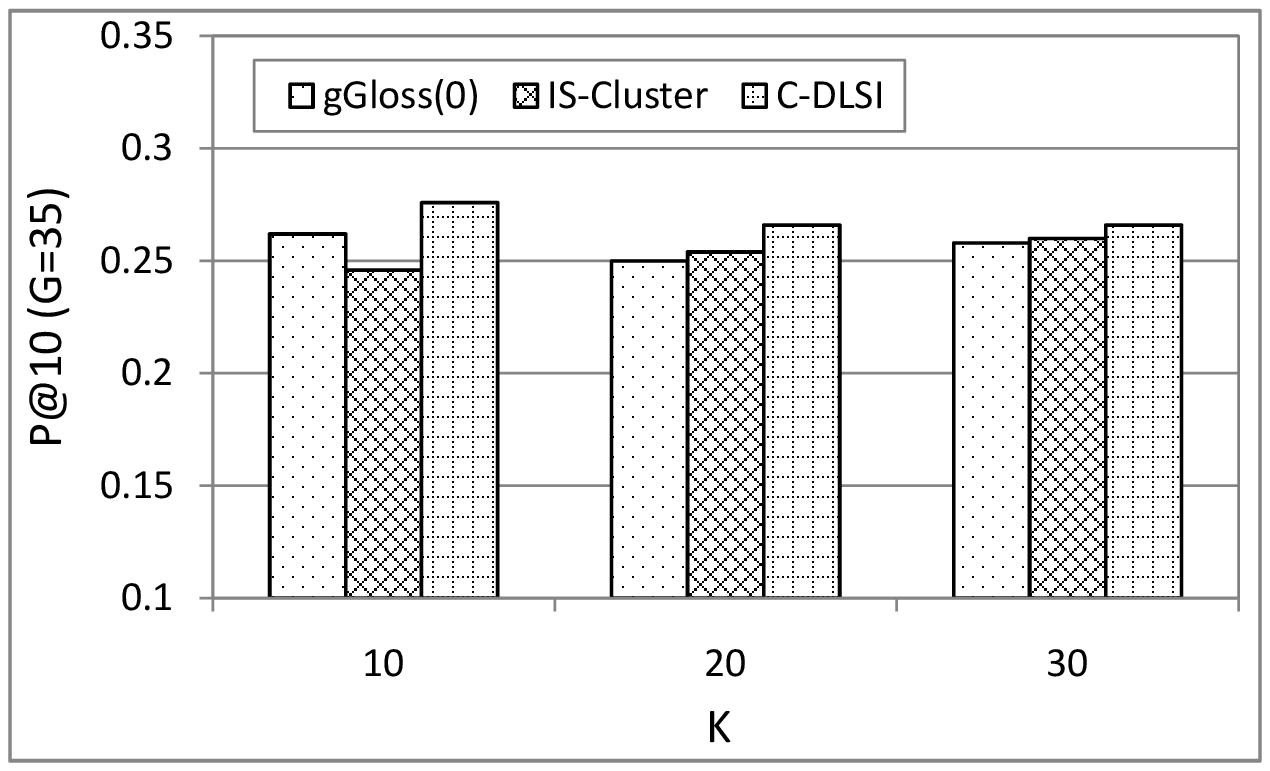, width=1.7in}
\caption{Performance of the three methods w. r. t. different cluster
numbers.}
\end{figure}

\subsubsection*{Cluster number $K$}
In general, the quality of k-means clustering is determined by the
preset cluster number $K$. To examine how much the FTR approaches
rely on the clustering quality, we investigate the performance of
the three methods with different cluster numbers, as presented in
Figure 11 (each cluster number corresponds to a different collection
assignment). The results show that IS-Cluster is more sensitive to
the cluster number. For larger cluster number $K$ (e.g., $K = 20$ or
$30$), IS-Cluster outperforms gGloss(0). However, for small cluster
number $K$ (e.g., $K = 10$), IS-Cluster may be beaten by gGloss(0).
It indicates that IS-Cluster relies more on the clustering quality
and we have to select proper $K$ to guarantee a good performance. On
the contrary, C-DLSI is more stable and substantially outperforms
gGloss(0) most of the time. This characteristic means a lot since
the broker usually inclines to keep a small directory for
scalability or bandwidth considerations. C-DLSI can better adapt to
systemns with limited resource.

\begin{figure}
\centering \epsfig{file=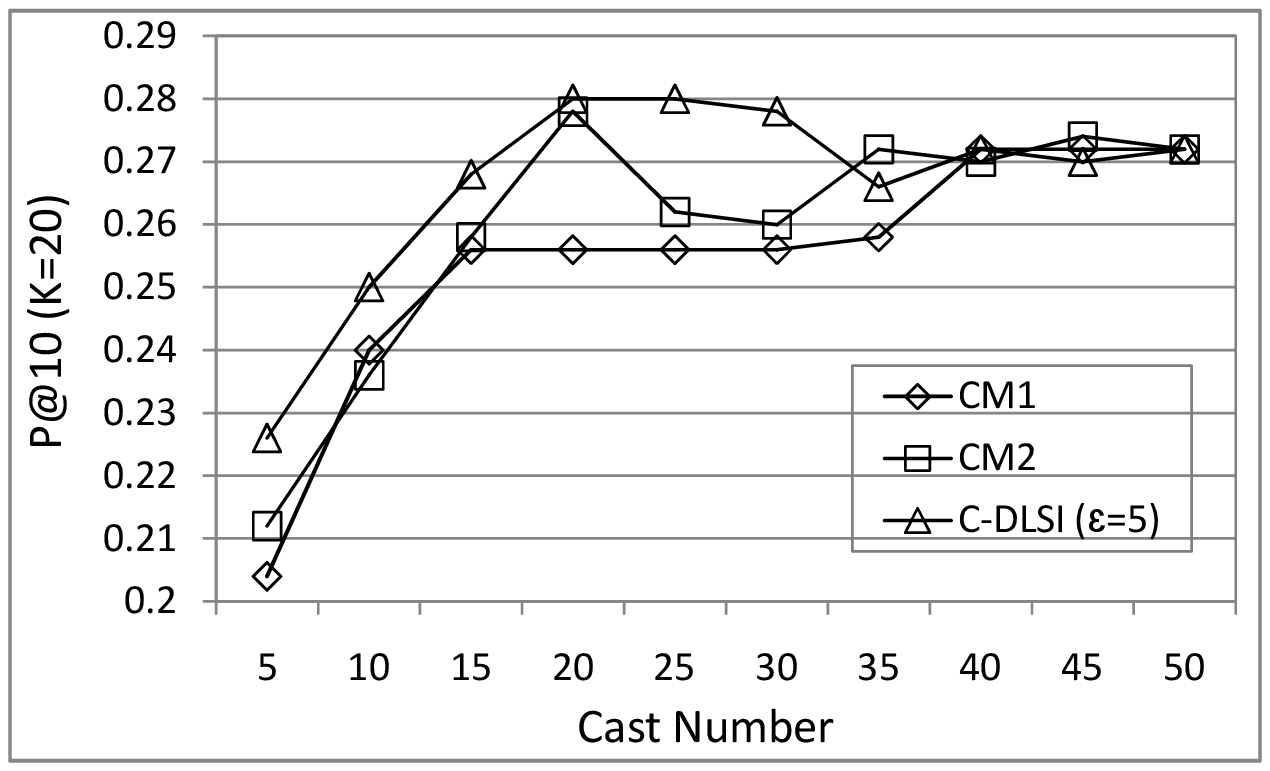, width=2.5in}
\caption{Performance comparison between C-DLSI and the combination
methods (IS-Cluster or gGloss(0) plus LSI-based IR). }
\end{figure}

\subsubsection{Compatibility Issue in FTR} As shown in Table 3,
C-DLSI exhibits very close performance to IS-Cluster on peer
selection. For some cast numbers, IS-cluster is even better. Based
on this, we may consider whether a combination of IS-Cluster and
LSI-based text retrieval is also a good choice compared to C-DLSI.
In this subsection, we examine two combination methods, namely CM1
and CM2. Both of them retrieve documents in each peer based on LSI
as in C-DLSI. However, for peer selection, CM1 uses gGloss(0) while
CM2 uses IS-Cluster. Figure 12 shows the result of this comparison
and it shows that C-DLSI still outperforms the combined methods. For
example, although IS-Cluster and gGloss(0) get slightly better peer
selection result than C-DLSI when cast number $T=25$, the
combination methods are still inferior to C-DLSI at $T=25$. From the
result, we can see that although IS-Cluster is able to detect the
semantic meaning of each document, it does not adapt to LSI in local
text retrieval (actually, it is difficult to find a proper local
retrieval scheme for IS-Cluster because of its special property as
we point out in Section 4.1), thus achieving little improvement.
Therefore, the performance of the combination method CM2 is only
slightly better than CM1. However, the peer selection method of
C-DLSI can distinguish each document and obtain proper semantic
spaces based on LSI. Therefore, it is more adaptive to the LSI-based
text retrieval.

\begin{figure}
\centering \epsfig{file=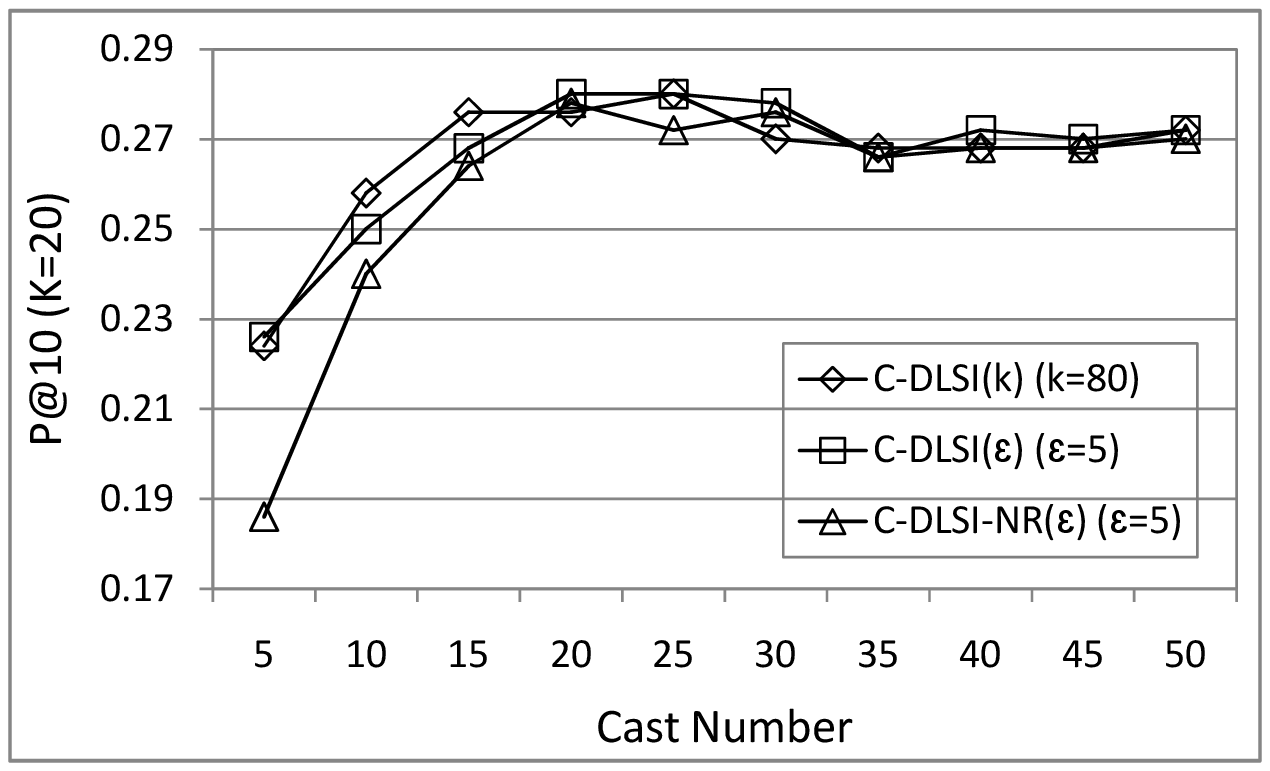, width=1.7in}
\epsfig{file=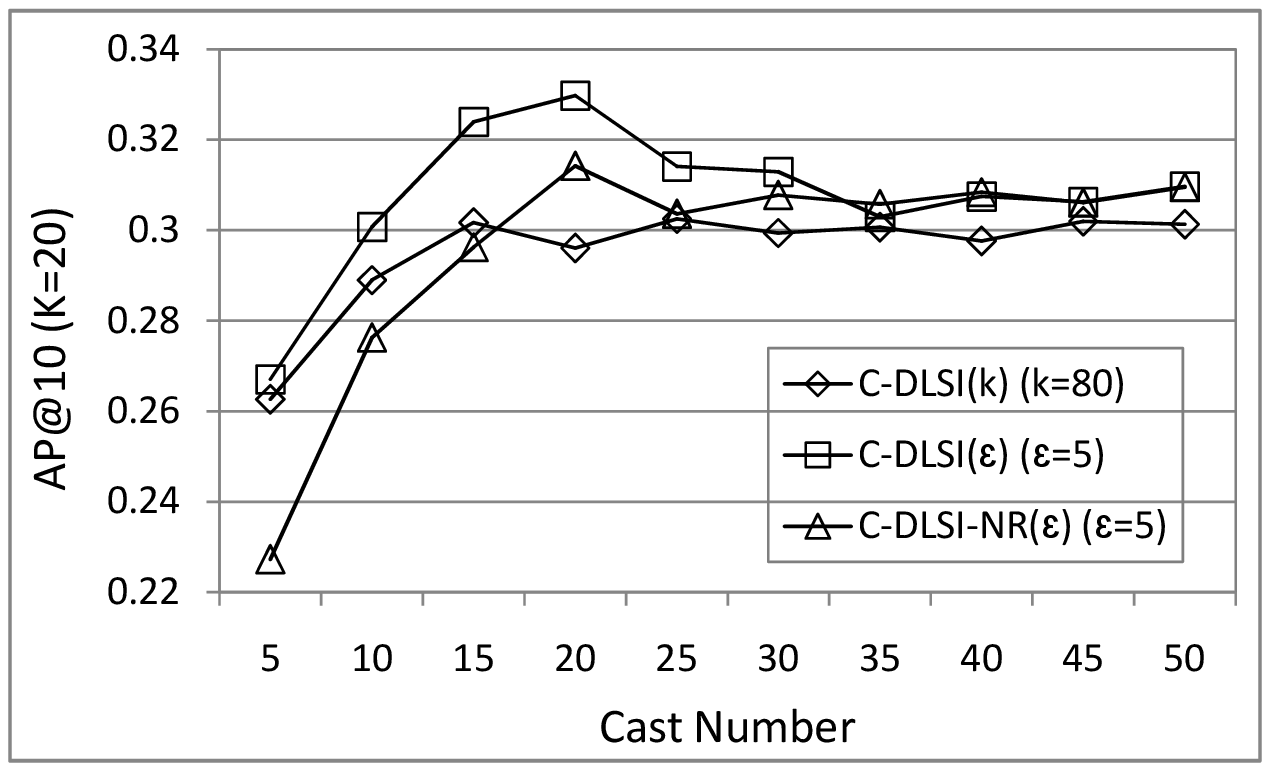, width=1.7in} \caption{Performance
comparison between C-DLSI($\varepsilon$) and C-DLSI($k$) for the
case $K=20$.}
\end{figure}

\begin{figure}
\centering \epsfig{file=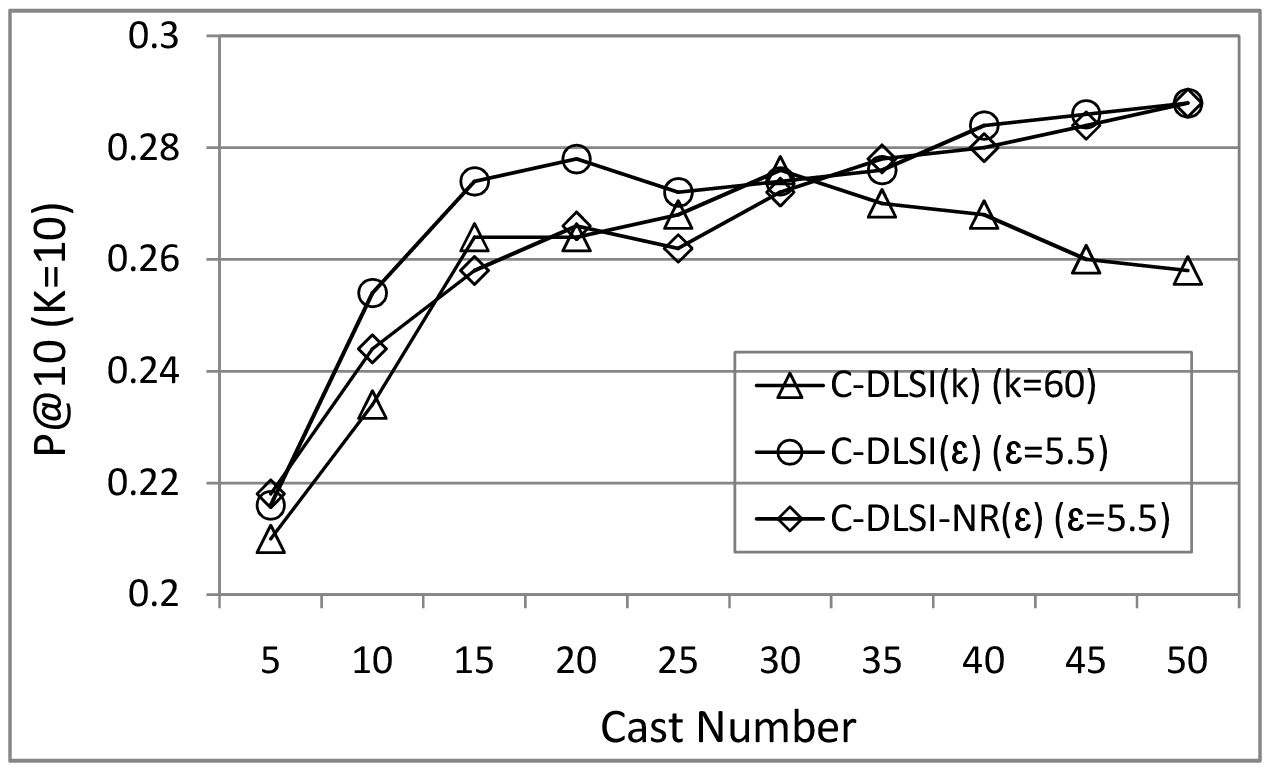, width=1.7in}
\epsfig{file=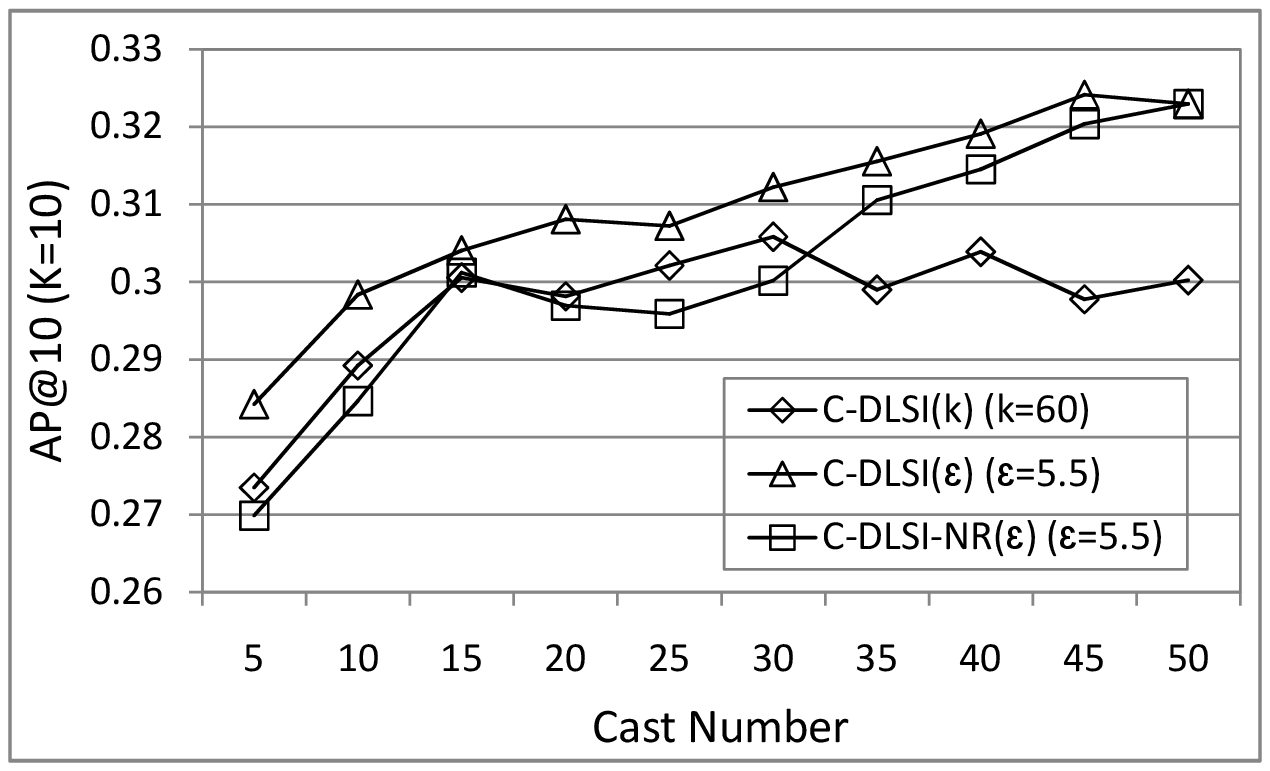, width=1.7in} \caption{Performance
comparison between C-DLSI($\varepsilon$) and C-DLSI($k$) for the
case $K=10$.}
\end{figure}

\begin{figure}
\centering \epsfig{file=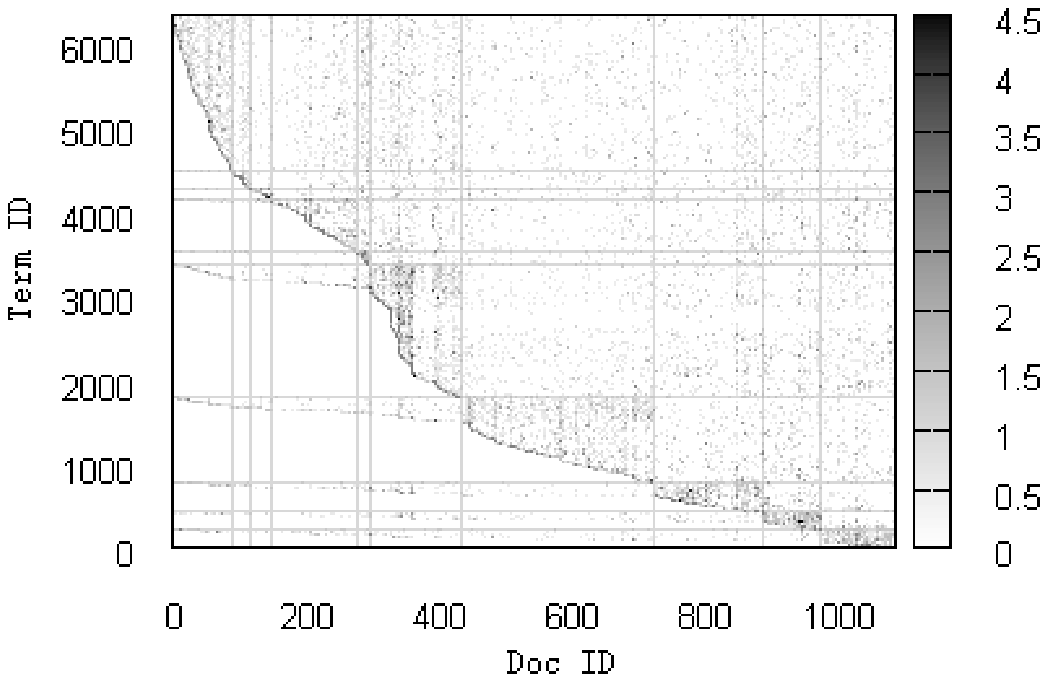, width=2.5in} \caption{An
example of gray-scale map for a peer when $K = 10$.}
\end{figure}

\subsubsection{C-DLSI($\varepsilon$) Vs. C-DLSI($k$)}
In this subsection, we compare our method C-DLSI, denoted as
C-DLSI($\varepsilon$), with another possible form of C-DLSI which is
based on a truncated number $k$, namely, C-DLSI($k$). Here we choose
the truncated value $k$ with the best performance of all possible
values for C-DLSI, e.g., $k = 80$ for the case $K=20$, and $k=60$
for the case $K=10$. The results are shown in Figure 13 and Figure
14 for the cases $K=20$ and $K=10$ respectively. We can see that
C-DLSI($\varepsilon$) in general beats C-DLSI($k$). In particular,
for the case $K=10$, when all of the indexing peers are selected ($G
= 50$), this gap becomes largest, indicating an $11.63\%$
improvement over C-DLSI($k$). We find that the numbers of the
relevant documents returned by the peers in two methods are quite
similar. It means C-DLSI($\varepsilon$) makes the LSI spaces among
clusters comparable and thus provides better result merging.
Finally, we also examine the C-DLSI scheme without considering the
cluster relations, denoted as C-DLSI-NR($\varepsilon$). The
performance are given in Figure 13 and Figure 14. Generally, by
considering cluster relations, C-DLSI($\varepsilon$) gains some
improvements compared to C-DLSI-NR($\varepsilon$). This gain becomes
larger for $K=20$, because the clustering quality in the case $K=20$
is worse than that of $K=10$ (as shown in Figure 15).

\subsubsection{Collection Update Scheme}
Finally, we will test our update scheme in FTR. In the experiment,
we only simulate one case of collection update, i.e., indexing new
documents. Specifically, we first index only $70\%$ of the total
documents and build the corresponding LSI. Then we gradually add new
documents from the remaining set, adding $5\%$ of the indexed size
in each step. Figure 16 shows an example $(K = 10, N = 10)$ of the
performance variation of the three methods during the update
procedure. We can see that the performance of C-DLSI still increases
with more documents added. For small cast number (e.g., $T = 10$),
C-DLSI outperforms (or at least be comparable to) gGloss(0)
(IS-Cluster) until the update amount reaches $30\%-35\%$
$(5\%-10\%)$. For larger cast number (e.g., $T = 20$), this valid
amount for C-DLSI before rebuilding the LSI decreases to $10\%-15\%$
($5\%-10\%$). Besides, we also get similar results for other
settings. It means that our update scheme is especially applicable
for FTR systems, in which the cast number is relatively small.

\begin{figure}
\centering \epsfig{file=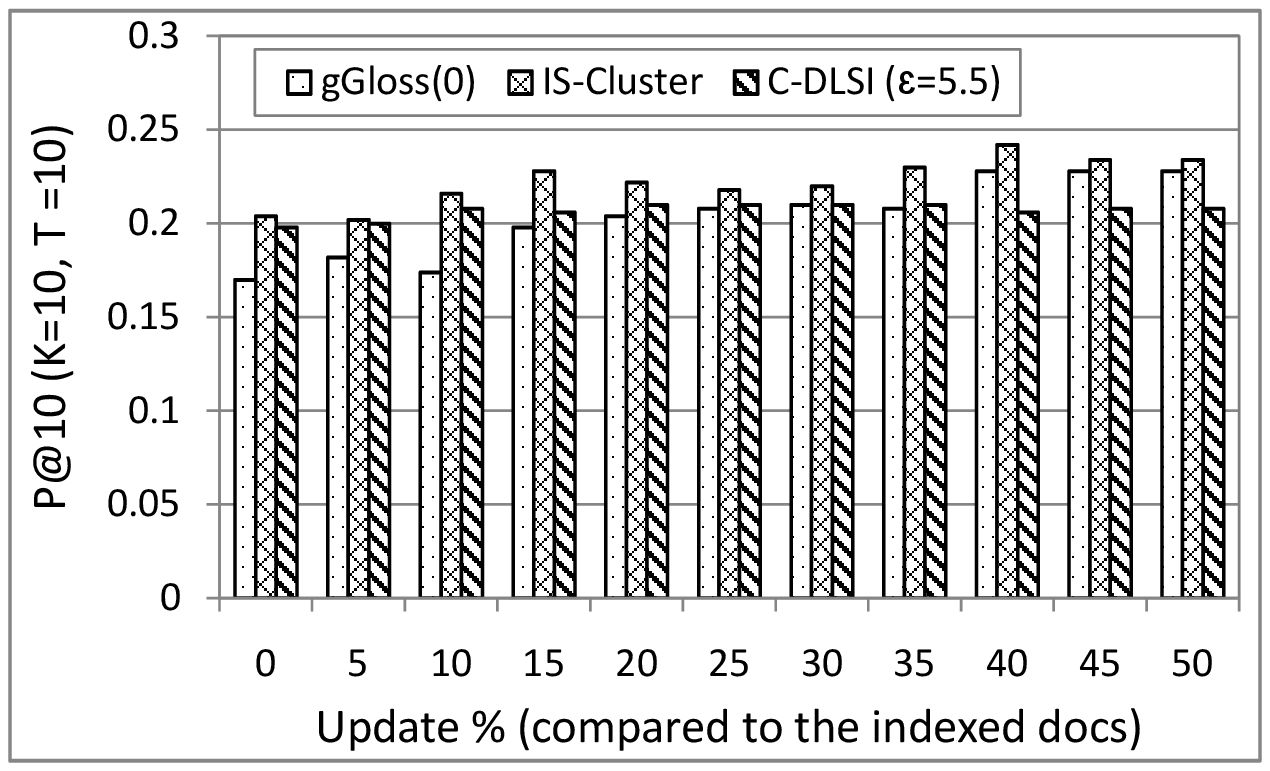, width=1.7in}
\epsfig{file=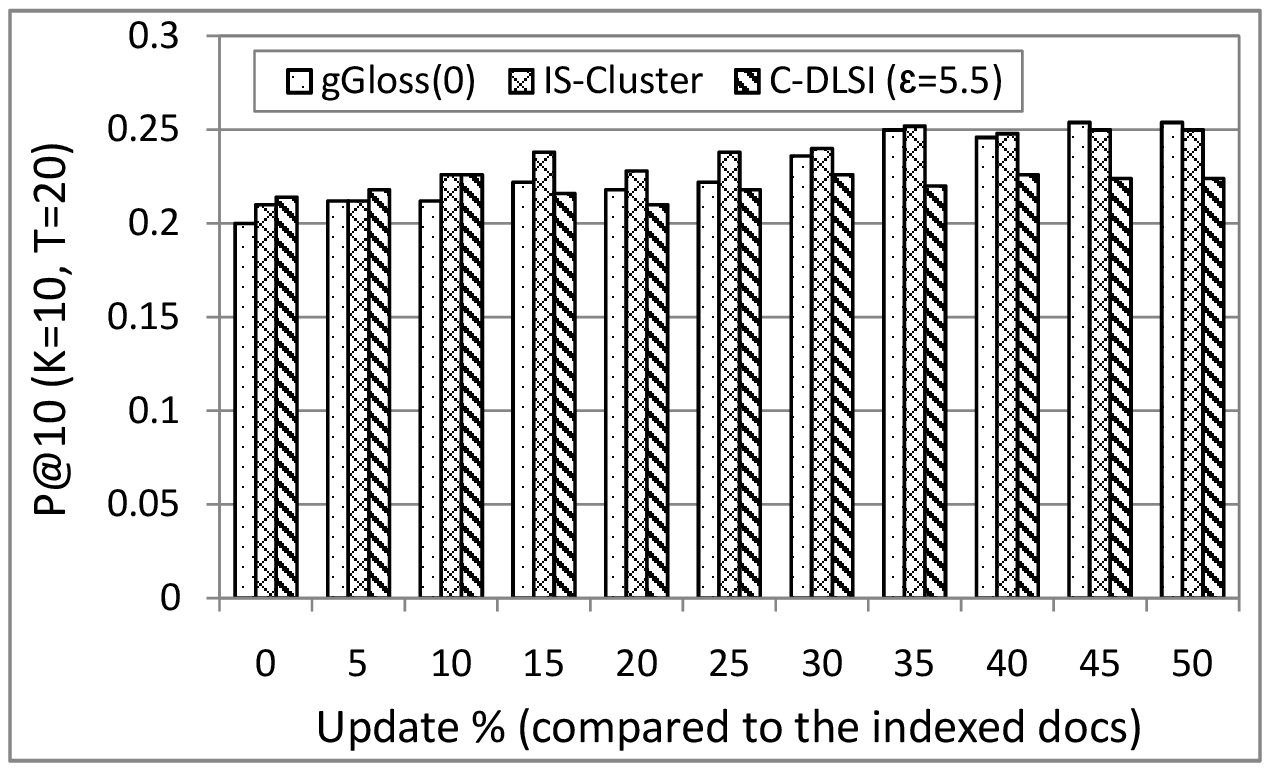, width=1.7in} \caption{Performance
comparison with increasing update. }
\end{figure}

\section{Conclusion}
In this paper, we proposed a promising solution for the challenges
of FTR. Different from the existing methods, our proposed method,
Cluster-based Distributed Latent Semantic Indexing (C-DLSI),
captures the semantic structure of a peer by identifying the LSI
spaces within the clusters and considering the relations among them,
thus providing more precise evaluation of the peer. We analyzed the
characteristics of C-DLSI, based on which novel descriptors of the
peers and the federated query processing was proposed. Besides, we
devised an effective form of C-DLSI, namely, C-DLSI($\varepsilon$),
the performance of which is studied and verified by using gray-scale
map in the experiments. Our method is efficient since only the
clusters affected by the updates need to be reindexed. Moreover, we
consider the update problem of C-DLSI and provide an update scheme
to make the framework more efficient while guaranteeing its
effectiveness. The experimental results confirmed the superiority of
our model and update scheme, and showed that our method outperforms
other existing methods including the previous cluster-based method.

\end{document}